\documentclass[a4paper,11pt]{article}

\pdfoutput=1 

\usepackage[T1]{fontenc} 
\usepackage{footnote}

\usepackage{graphicx,ulem}
\usepackage{longtable}
\usepackage{float}
\usepackage{dcolumn}
\usepackage{graphics,epsfig}
\usepackage{amsmath,amssymb,latexsym,mathrsfs}
\usepackage{bm}
\usepackage{color}
\usepackage{color}
\usepackage{subfigure}
\usepackage{multirow}
\usepackage{amsfonts}

\usepackage[margin=1.0in]{geometry}
\usepackage{graphicx}
\usepackage{afterpage}
\usepackage{float}
\usepackage{amsmath}
\usepackage[T1]{fontenc}

\usepackage{tikz}
\usetikzlibrary{arrows,arrows.meta}
\usetikzlibrary{decorations.markings}
\usetikzlibrary{decorations.pathmorphing}
\usetikzlibrary{positioning,shapes,backgrounds}
\usetikzlibrary{calc,intersections,patterns,fit,automata,decorations}
\usepackage{tikzsymbols}
\usepackage{paralist}
\usepackage{enumitem}

\newcommand{\remove}[1]{}

\definecolor{skipcolor}{rgb}{1, .92, .92}
\definecolor{rejoincolor}{rgb}{.9, 1, 0.9}
\definecolor{frameskip}{rgb}{0.95, 0, 0}
\definecolor{framerejoin}{rgb}{0, .75, 0}

\definecolor{gold}{rgb}{1,0.553,0}
\definecolor{lightbrown}{rgb}{0.9305,0.86275,0.8}
\definecolor{fillcopper}{rgb}{0.722,0.451,0.2}
\definecolor{fillsilver}{rgb}{0.753,0.753,0.753}
\definecolor{fillgray}{rgb}{0.4,0.4,0.4}
\definecolor{filllightgray}{rgb}{0.65,0.65,0.65}
\definecolor{fillLightgray}{rgb}{0.85,0.85,0.85}
\definecolor{fillred}{rgb}{1,0.2,0.2}
\definecolor{filllightred}{rgb}{1,0.6,0.6}
\definecolor{fillblue}{rgb}{0.2,0.35,1}
\definecolor{filllightblue}{rgb}{0.85,0.85,1}
\definecolor{fillgreen}{rgb}{0.2,0.7,0.2}
\definecolor{filllightgreen}{rgb}{0.9,1,0.9}
\definecolor{fillgreenbox}{rgb}{0.9,1,0.9}
\definecolor{fillpurple}{rgb}{0.7,0.4,0.74}

\def\math#1{$#1$}

\def\mymathskip{4.5pt}

\def\mldc#1{\mld{\abovedisplayskip=\mymathskip plus 1pt minus 1pt%
\abovedisplayshortskip=0pt plus 1pt minus 1pt%
\belowdisplayskip=\mymathskip plus 1pt minus 1pt%
\belowdisplayshortskip=0pt plus 1pt minus 1pt%
#1}}

\def\mld#1{\begin{equation}
#1
\end{equation}}

\def\eqar#1{\begin{eqnarray}
#1
\end{eqnarray}}

\def\frac#1#2{{#1\over #2}}

\DeclareSymbolFont{AMSb}{U}{msb}{m}{n}

\def\argmin{\mathop{\hbox{argmin}}\limits}

\def\r#1{{\eqref{#1}}}

\newcounter{rmnum}
\def\RN#1{\setcounter{rmnum}{#1}\uppercase\expandafter{\romannumeral\value{rmnum}}}
\def\rn#1{\setcounter{rmnum}{#1}\expandafter{\romannumeral\value{rmnum}}}

\remove{

}

\definecolor{shadecolor}{gray}{.85}%
\definecolor{tintedcolor}{gray}{.8}%

{\makeatletter\gdef\reallynopagebreak{\par\nopagebreak\@nobreaktrue}}

\providecommand\remove[1]{}

\DeclareSymbolFont{extraup}{U}{zavm}{m}{n}
\DeclareMathSymbol{\varheart}{\mathalpha}{extraup}{86}
\DeclareMathSymbol{\vardiamond}{\mathalpha}{extraup}{87}

\author{Liam Dowling Jones$^1$, Malik Magdon-Ismail$^2$, Laura Mersini-Houghton$^1$ and Steven Meshnick$^3$}

\begin{document}


\title{A New Mathematical Model for Controlled Pandemics Like COVID-19 : AI Implemented Predictions}

\maketitle

\newenvironment{affiliations}{%
    \setcounter{enumi}{1}%
    \setlength{\parindent}{0in}%
    \slshape\sloppy%
    \begin{list}{\upshape$^{\arabic{enumi}}$}{%
        \usecounter{enumi}%
        \setlength{\leftmargin}{0in}%
        \setlength{\topsep}{0in}%
        \setlength{\labelsep}{0in}%
        \setlength{\labelwidth}{0in}%
        \setlength{\listparindent}{0in}%
        \setlength{\itemsep}{0ex}%
        \setlength{\parsep}{0in}%
        }
    }{\end{list}\par\vspace{11pt}}

\begin{affiliations}
 \item Department of Physics and Astronomy and COSMS Institute, UNC- Chapel Hill, NC 27599, USA
 \item Computer Science Department, Rensselauer Polytechnic Institute,110 8th Street, Troy, NY 12180, USA
 \item UNC Gillings School of Global Public Health, UNC-Chapel Hill, NC 27599-7435, USA 
\end{affiliations}



\begin{abstract}

We present a new mathematical model to explicitly capture the effects that the three restriction measures: the lockdown date and duration, social distancing and masks, and, schools and border closing, have in controlling the spread of COVID-19 infections $i(r, t)$. Before restrictions were introduced, the random spread of infections as described by the SEIR model grew exponentially. The addition of control measures introduces a mixing of order and disorder in the system's evolution which fall under a different mathematical class of models that can eventually lead to critical phenomena. 
A generic analytical solution is hard to obtain. We use machine learning to solve the new equations for $i(r,t)$, the infections $i$ in any region $r$ at time $t$ and derive predictions for the spread of infections over time as a function of the strength of the specific measure taken and their duration. The machine is trained in all of the COVID-19 published data for each region, county, state, and country in the world. It utilizes optimization to learn the best-fit values of the model's parameters from past data in each region in the world, and it updates the predicted infections curves for any future restrictions that may be added or relaxed anywhere. 
We hope this interdisciplinary effort, a new mathematical model that predicts the impact of each measure in slowing down infection spread combined with the solving power of machine learning, is a useful tool in the fight against the current pandemic and potentially future ones.
    
\end{abstract}


\raggedbottom

\section{Introduction}

It is the first time in recent human history where the spread of a novel virus is not allowed to be random but strict control measures like: locking down countries, closing borders, masks and social distancing, closing schools, and quarantining are undertaken at such a {\it massive global scale}. New measures require a new mathematical model which transcends the traditional $SEIR$ model used in epidemiology to study the spread of viral diseases: a model which will allow us to predict and prepare for pandemics well in advance by enabling us to evaluate the combination of restrictions with the highest impact.

When a virus diffuses unrestricted through a population, it spreads randomly. Therefore, the number of infections grows exponentially with time, obeying a nearly Gaussian distribution curve, the hallmark of randomness. The distribution of random infections in each region before control measures are introduced, map human social behavior (contact rate) as well as population density and demographics. The epidemiological model which has been used to describe the spreading of viruses through a population with $N$ people is the class of $SEIR$ models. The driving force of infections in the $SEIR$ type models is a coupling between the susceptibles and infected groups where their coupling strength $\beta$ is directly proportional to the now famous contagion parameter $R_0$.

In this work we propose a stochastic mathematical model for the infection spread which, in addition to the driving force of infections $\beta$ term of the $SEIR$ model, explicitly adds the impact of each of the three measures (lockdown, social distancing plus masks, and school and border closing) imposed worldwide in disrupting exposure and breaking the transmission of the pandemic have on the diffusion of infection through population. Since time is of essence during a pandemic, we implement our new model and solve its equations using an AI approach to predict what will happen for the rest of the year in any region. The machine is trained in all the $Covid-19$ data available from Johns Hopkins University for any region, county, state, or country before and after restrictions were imposed. 

Data from the early phase of the pandemic when the spread of infection was unrestricted is very useful in training the machine to learn, and include in its estimate for the restrictions parameters introduced in our model, the population distribution and demographics as well as the characteristic social behavior (contact rates) of each region. This advanced approach allows us to estimate and predict in a matter of seconds how the infections may spread for the rest of the year in any region as a function of the time and type of control measures being introduced and their population characteristics. 

Furthermore, we offer an estimate of the number of infection cases undetected for each confirmed positive test in a region, a parameter we call $\gamma (\vec{r}, t)$. Through this parameter we can convert the confirmed daily infection curves for any region into a knowledge of the 'reservoir', that is, the total number of the infected population in that region at any time whether they are tested or not. The machine also learns to account for the time delay between infection and its reported tested confirmation. However, since the machine is trained on historic public data, the findings reported here are limited by the accuracy of the reported data: the time lapse between the true first case of infection in a region and the first confirmed positive test, as well as the noise in the reported data, i.e. the amount of uncertainty in the public data for confirmed positives.

Let us first review the basics and terminology of the $SEIR$ model before replacing it with a new model which adds new terms to include the restrictions measures.
The letters in $SEIR$ stand for 'number of people susceptible $S$ to the virus, number of people exposed $E$ to it, the number of infected people $I$, and the number of people removed or recovered $\cal{R}$', within that population of size $N$. In its simplest form it is mathematically described by \cite{imperial,epidemiology}:

\begin{equation} \left\{
\begin{array}{@{}rl@{}}
 \medskip
 \dfrac{ds}{dt} = -\beta si,~ s= S/N\\
 \medskip
 \dfrac{di}{dt} = \beta si -\tilde{\nu} i, ~ i= I/N\\
 \dfrac{dr}{dr} = \tilde{\nu} i, ~ r= \cal{R}/N\\
\end{array}
\right .
\label{seireqns}
\end{equation}

The parameters $(\tilde{\nu} , \beta)$, the removal rate and effective contact rate parameters respectively, are in general time dependent.
Without loosing generality, the exposed $e = E/N$ and infected $i= I/N$ percentages of the population in the above equations are grouped together into $i$ for simplicity. An explicit equation for $E$ mathematically obeys the same first order differential equation sourced by the susceptible population on the right hand side.

Combining the above and recalling that $e$ is included in $i$, gives:

\begin{equation}
\dfrac{d(i + r)}{dt} = \beta s i = - \frac{ds}{dt}
\label{infectioneqn}
\end{equation}

Not surprisingly, as can be seen from the opposite signs in the time derivative terms of $s$ and $(i+r)$ in Eqn.\ref{infectioneqn} what is a driving force for infections, $F_{seir} = \beta s i $, is a damping force with the same strength for the susceptible population. We will make use of Eqn.\ref{infectioneqn} in our model below.
 
Traditionally, Eqns. \ref{seireqns} are combined in a different way, in the manner of Eqn. \ref{infectRo} below:

\begin{equation}
\frac{di}{ds} = -1 + \frac{\tilde{\nu}}{\beta s} = -1 + \frac{1}{R_{0} s}, 
\label{infectRo}
\end{equation}

because of the explicit dependence on the crucial parameter $R_{0}$, the viral reproduction number, defined as

\begin{equation}
R_{o} = \frac{\beta}{\tilde{\nu}} = \tau \tilde{c}\frac{1}{\tilde{\nu}}  
\label{Roeqn}
\end{equation}

where the terms on the right hand side are respectively':

\begin{equation}
\tau = \text{Transmissibility given contact of \textit{s} with \textit{i}},
\end{equation}

\begin{equation}
\tilde{c} = \text{rate of contact between \textit{i} and \textit{s}},
\end{equation}

and, 

\begin{equation}
T = \frac{1}{\tilde{\nu}} = \text{duration of infection}
\end{equation}

It is clear from Eqn.\ref{Roeqn} and Eqn. \ref{infectRo} that $R_o$ describes how infectious a virus is since it gives the average number of people who will get the disease from one infected person. As such its value is crucial in estimating how fast the infection can spread uncontrolled. It should also be noted that in the early stages when the virus spreads unrestricted, this parameter depends on the population density and social behaviour of the region. We report $R_0$ dependence on a region's characteristics for three such examples in Section 4. For the case of COVID-19 epidemiological studies indicate that when the virus is spreading unrestricted at $t=0$, $R_{0}(t=0)$ can range from $2.4$ \cite{imperial} to $5.7$ \cite{Rocdcstudy}. Additionally, $R_0$ is an effective parameter since it changes over time. Integrating Eqn.\ref{infectRo} highlights the well known fact that damping the spread of infections translates into bringing $R_{0}$ to stable values of one or less. Mathematically, $\beta(t) = \tilde{\nu} R_{0}(t)$ contains the same information as $R_0$ on how the spread of the disease changes over time because $\tilde{\nu}$ is on average \footnote{Perhaps different strains of the virus may have different removal rates or their mutation may change $\tilde{\nu}$. }an unchanging characteristic of the virus. Therefore, any changes introduced by restrictions on the effective reproduction number $R_0$ of a region over time translate into changes in $\beta(t)$. 
    
The $SEIR$ model has been further compartmentalized in a $S E_{j}...I_{j}...R$ by breaking down the number of exposed  ``$E_{i}$'' and infected  ``$I_{j}$'' cases in sub-classes of the different stages of exposure, infection and quarantine, as studied in \cite{harvard, imperial, ihme, epidemiology}). Understanding these compartments is important for epidemiological contact tracing and surveillance and for clinical purposes when studying the probability of infection of a region. However, we won't need the details of the compartmentalized $SEIR$ models in what follows, as their net effect is absorbed in the $\beta$ driving force term already embedded in the new model below, and implicitly contained in the data on which the machine is trained and which it uses to optimize the best fit for $\beta(t)$ in real time.
  
The paper is organized as follows: In Section 2 we introduce the new mathematical model and describe the long term behavior of the model. In Section 3 we describe the method for obtaining the solutions using machine learning. In Section 4 we present the results of our AI implemented model and illustrate them for the case of three US states by showing the predicted curves of infection growth as a function of a representative combination of control measures given by their three parameters. We also estimate and show the number of true infections, the $reservoir$, for each confirmed infection, a number which for example in the US varies from 4 to 11 depending on the state. We are making the mathematical model and the code publicly available here \cite{ourcode}. This will allow users to instantly derive how the predicted growth of infections changes in their region any time a measure or a combination of them may be introduced and of the time chosen to impose those measures in the future. 
We summarize our findings and their usefulness for public health in Section 5.

We hope that the interdisciplinary approach and the tools offered here, that is, a combination of a realistic mathematical model with the speed of AI implemented predictions in the code provided will be useful in epidemiology, economics and public health considerations.

\section{The Mathematical Model For A Controlled Pandemic As A Universal Critical Phenomenon}

To account for the order introduced by the new global measures in slowing the random virus spreading anywhere in space and time, we enhance the $SEIR$ type models by adding new force terms to Eqns. \ref{seireqns} which capture the damping forces of infections imposed by lockdowns, masks and social distancing, and schools and border closings. The remaining random source of infection spread, not contained in the above, is given by a white noise and captured by a diffusion parameter. These terms are in addition to the $\beta s i$ term of Eqn. \ref{infectioneqn}, the driving force of infections which describes the contagion spread through random contact rate between the infected and susceptible populations in the $SEIR$ model. 

As described below, the addition of the new forces to Eqn. \ref{infectioneqn} leads to a new stochastic model which is in a completely different mathematical class from the $SEIR$ models. Let's promote the nonsusceptibles, i.e. the part of the population exposed to the virus at any level of infection, into an 'infection field': $\tilde{i}(\vec{x}, t) = i(\vec{x}, t) + e(\vec{x}, t) + r(\vec{x}, t)$ \footnote{We changed notation for spatial coordinates from $\vec{r}$ to $\vec{x}$ here to avoid any confusion of the SEIR variable $r$ for the recovered population with the spatial coordinate $\vec{r}$. However both $\vec{r}$ and $\vec{x}$ have the same meaning, region's location on a two dimensional surface and are used interchangeably. } spreading on $(2+1)-D$ (two spatial dimensions $\vec{x}$ and one temporal dimension $t$).

In this model we make the following {\it assumptions}: ${\it a)}$ no vaccine will become massively available until next year; ${\it b)}$ the recovered, infected and exposed (the $EIR$ part in $SEIR$) will gain some immunity, therefore will not be susceptible to being sick again this year or until the vaccine is available. The latter is expected on general grounds on how our immunity system protects us from viruses, but it is not verified to be true for the novel COVID-19 virus. (Indeed, recent medical findings \cite{reinfection} seem to indicate COVID-19 may be unique in the sense that those who recovered from it may not have a long lasting immunity to reinfection.)  

The infection field \math{\tilde{{i}}(\vec{x}, t)} is a restricted self-avoiding random walker on a two dimensional space and in time. 
The {\it restricted} feature on the random walk of the infection seeping through the population of some region of space $\vec{x}$ with density $\rho(\vec{x}, t)$ at some time $t$ is due to the control measures taken in restricting the virus spreading. The {\it self-avoiding} feature is related to the assumption of immunity, namely a person who has already been infected once will not be infected again during the time interval from now to until a vaccine is available. Typically, this class of models is expected to lead to critical phenomena of polynomial growth of infections at large times which display a universal scale-free behavior described by what in critical phenomena are known as critical exponents $\nu, \mu$ \footnote{The critical exponent $\nu$ here should not be confused with the removal rate parameter $\tilde{\nu}$ in Eqns. \ref{seireqns} of the SEIR model}. In two spatial dimensions $\nu$ provides the fractal critical dimension of the fractal space available to the restricted random walker \math{\tilde{{i}}(\vec{x}, t)} for that region. According to Flory's theory \cite{kardar, kardar2, flory} for critical exponents, analytically the fractal dimension $\nu$ in two dimensions is expected to be about $2 \nu\sim 3/2$ instead of $2$. That is, without vaccines at large times \math{\tilde{{i}}(\vec{x}, t)} asymptotically is expected to approach a stable fractal distribution of self similar 'hot' and 'cold' infection clusters on a space of dimension less than $2$.  Despite a vast literature on the subject, analytical results in a closed form for this class of critical phenomena for a restricted self avoiding (SAW) random walker in two spatial dimensions do not exist even for the simpler case of time independent parameters. Our model therefore has an additional layer of complexity to the mathematical results cited here for a two dimensional restricted SAW random walker: the restrictions on our random walker \math{\tilde{{i}}(\vec{x}, t)} are time dependent. Therefore we will use machine learning for solving these equations to predict the late time growth of infection spreading as a function of restriction measures.

\subsection{The Mathematical Model For Restricted Infection Spread}

Consider the infection variable $\tilde{i}$ to be  {\it our system}. We promoted it into a field $i(\vec{x},t)$ random walking through a 'host' environment field $s(\vec{x}, t)$ living on a two dimensional region over time. In what follows, we will also make use of the parameter $\gamma$ mentioned in Section 1, defined in Eqn.\ref{gamma} as the ratio between the {\it confirmed} infections $i(\vec{x}, t)$ and the actual reservoir of the non-susceptible population $\tilde{i}(\vec{x}, t)$. 

\begin{equation}
\gamma(\vec{x}, t) = \frac{i(\vec{x}, t)}{\tilde{i}(\vec{x}, t)},
\label{gamma}
\end{equation}

Clearly, $\gamma(\vec{x}, t)$ is a space and time dependent parameter since it quantifies the true number of infections in a population located at $\vec{x}$ per each confirmed case there by testing, at any time $t$.

{\it The Force of Lockdown:} 

Let us now model the effect of the lockdown measure on reducing the infection spread with a quadratic confining potential $V_{i} (\tilde{i})$. A single well potential for a field, in our case the active infections, describes the effect of lockdown in a region located at $\vec{r}$ because such a potential drives the active confirmed infection field $i(\vec{r}, t)$ to eventually rolls down to zero. \footnote{A quartic or any type of concave single well confining potential function would work equally well as the quadratic type of Eqn.\ref{lockdownV}, so we don't loose any information by choosing the simplest such potential.}  The strength of the confinement potential camptures the impact of the lockdown in damping active infections and is given by its space and time dependent coupling constant $\alpha = \alpha[(t_{f} - t_{i}), t, r]$, which varies with region $\vec{r}$ and depends on $t_{i}, t_{f}$, the initial and final date (duration) of the lockdown of the region, respectively. Since our system is $\tilde{i}(\vec{r}, t)$ then, using the rescaling between confirmed infection $i$ and $\tilde{i}$ of Eqn. \ref{gamma}, we can rewrite the lockdown potential in restricting the random walker in terms of $\tilde{i}$ by absorbing $\gamma(\vec{r}, t)$ into the strength $\alpha(\vec{r}, t)$ of the lockdown potential $V_{i}$.

With these considerations the lockdown potential becomes

\begin{equation}
V_{i} = \dfrac{\tilde{\alpha}~\tilde{i}^{2}}{2}
\label{lockdownV}
\end{equation}

where the rescaled lockdown strength is: $\tilde{\alpha} = \frac{\alpha}{\gamma^{2}}$. The damping force of the lockdown on infections is simply the derivative of its potential $F_{i} = -\frac{d V_{i}}{d \tilde{i}}$. From now we will suppress showing the symbols $t_{i}, t_{f}$ in $\alpha$ for ease of notation.

{\it The Force of Social Distancing and Masks:}

The purpose of masks and social distancing is to eliminate the possibility of exposure to the virus, in other words to minimize the chance that susceptibles are physically close to the path of droplets from others, or to block their flux with a mask altogether. We here assume that the part of the population which follows guidelines for eliminating exposure to the virus is equally likely to engage both methods to block exposure, masks and social distancing, therefore the impact of the social distancing potential in slowing down infection spread discussed here includs the effect of masks as well. While we don't include any of the biological or clinical parameters such as the role of viral load in infecting susceptibles, the risk from potential airborne properties, or a variation in the size and trajectory of droplets, we gauge the net effect of social distancing in eliminating exposure to droplets by $i(\vec{r}, t)$ as a damping force on infections growth by relying on a simple physics based approximation: the trajectory of a flux of droplets ejected by $i$ (through breathing, coughing and sneezing) follows a typical classical projectile motion described by Newtonian gravity. While estimating the projectile distance a droplet may travel before hitting the ground as a function of its weight, size and ejection speed is not hard, such details are not important on the solution and stochastic evolution of our system. Therefore in what follows, we ignore droplet size effects. We average over the size and initial speed of droplets and consider them collectively as a flux of particles moving together under the gravitational force, along the same projectile trajectory.    

Normally, the 'contagion interaction' between a $j^{th}$ susceptible person $s_{j}$ and droplets flux from a $k^{th}$ infected person $i_{k}$ would take the form $V_{si} = - \tilde{\delta} \Sigma \frac{s_{j} i_{k}}{r_{j k}^{\zeta}}$, where $\tilde{\delta}$ is the strength parameter of the potential, $r_{j k}$ is the distance between them, and the sum is over all $(j, k)$. For the purely gravitational forces on the droplets motion $\zeta =2$. For a more detailed analysis which may account for changes in the projectile motion due to other forces on the droplets such a drifting force from wind and other weather elements, or an air buoyancy force on the droplets inducing floating, we could allow $\zeta$ to be an additional parameter $\zeta \ne 2$ the value of each would be determined by simulations. Since the susceptible part of the population is much larger than the infected part so far in all the reported data for any region, we replace the sum over j of $s_{j}$ with an averaged value $s(\vec{r}, t) = (1 - \tilde{i})$, up to a proportionality constant which depends on the population distribution and social behavior absorbed in $\tilde{\delta}$ for each region, through machine learning. As before, the re-scaling factor $\gamma$ between $i$ and $\tilde{i}$ of Eqn.\ref{gamma} is included in $\tilde{\delta}$. As mentioned the machine is trained to learn and include the information about the population density distribution $\rho(\vec{r}, t)$ and social interactions from earlier times data when the spread was random of each region within radius $\vec{r}$ at time $t$ into the coupling constants $\alpha, \tilde{\delta}$ of each potential. This allows us to further approximate the typical average distance $r \approx <r_{j k}>$ between infected and susceptible people from different households on a two dimensional surface with an ensemble averaged distance $r \approx \tilde{i}^{-1/2}$ for that region's population density.Any numerical factors that arise from this approximation are roughly \math{\tilde{i}^(-1/2)} and are also absorbed by \math{\tilde{\delta}}. Therefore, we denote the rescaled strength parameter of this potential by $\delta $. Under the above simplifications, the potential $V_D$ capturing the effect of social distancing and masks restriction designed to avoid contact between droplets from the infected and susceptibles of $V_{s, i}$ interactions takes the following form

\begin{equation}
V_{D} = \dfrac{ \delta ~ \tilde{i}~ s}{r^{\zeta}}
\label{socialdistanceV}
\end{equation}

where 
\begin{equation}
s = (1 - \tilde{i})
\end{equation}

To reduce the number of parameters the machine needs to estimate and best fit to data, we now take $\zeta$ = 2.
The damping force on infections from the masks and social distancing potential, then is $F_{D} = -\frac{d V_{D}}{d \tilde{i}}$. Similarly the coupling strength $\beta$ for the driving force of infection spread depends on space and time.

Collecting all the restriction forces acting on the infection field $\tilde{i}$ leads to

\begin{equation}
\frac{d\tilde{i}}{dt} \approx \tilde{\beta}(1-i)\tilde{i} - \tilde{\alpha}\tilde{i} - \dfrac{\delta (1 -2\tilde{i})}{r^{2}} + f(t) = \tilde{\beta} ~ s ~ \tilde{i} - V'_{i} - V'_{D} + f(t)
\label{rateofinfection}
\end{equation}

The prime denotes $\frac{d}{d \tilde{i}}$. The first term in Eqn.\ref{rateofinfection} is the familiar $SEIR$ driving force of infections, which in our model presented here corresponds to a potential function, $V_{SEIR} = - \beta \frac{s i^{2}}{2}$. \math{\tilde{\beta}} is the space and time dependent coefficient of \math{\beta} which is rescaled by \math{\gamma} and by population demographics and social behavior; the second term is the damping force with finite duration due to the lockdown; the third term is the damping force on infections from the percent of population which follow rules on masks and social distancing. 

{\it The Noise Term from the Environment and Diffusion:}

The last term $f(t)$ in Eqn.\ref{rateofinfection} is an important one which deserves some explanation. This term is a source of noise from the background $s(\vec{r}, t)$ acting on the system $\tilde{i}$, a random local force driving infection spreading details of which are not known. The only information known about the noise term $f(t)$ is its averaged values, such that on average $<f(t)> = 0$ and $< f(t) f(t') > = D \delta(t - t')$, where $D$ is the diffusion parameter. The noise from the environment $f(t)$ onto our system gauges the fact that for as long as there are hosts $s(r, t)$ on which the virus can feed, then some random diffusion of the infection seeping through a population may still persist and escape restrictions or detection despite efforts carried out to control it. This term quantifies what percentage of the population, may not or can not follow the control measures or any accidental exposure to the virus despite best efforts. 

With the noise term $f(t)$ included, Eqn. \ref{rateofinfection} becomes a stochastic differential equation known as the Langevin equation \cite{fokkerplanck}

Given $\tilde{i}= e + i + r$ and its relation to $i$ through the scaling factor $\gamma$, from Eqn. \ref{rateofinfection} we could also rewrite the Langevin equation \ref{rateofinfection} as an equation for the rate of decrease of the 'hosts' environment, the susceptible population

\begin{equation}
\frac{ds}{dt} = -[\tilde{\beta} ~ s ~ \tilde{i} - V'_{i} - V'_{D} + f(t)]= -[V'_{total} + f(t)].
\label{rateofhealthy}
\end{equation}

Eqn.\ref{rateofhealthy} relates how decreasing the ``driving'' force of infection through these measures slows down the ``damping'' rate of the susceptibles background. \footnote{The relation $\frac{d s}{d t} = - \frac{d \tilde{i}}{d t}$ between the susceptible background $s$ and the infection random walker $\tilde{i}$ for the stationary case eventually leads to the distribution of 'hot' and 'cold' clusters of infection in the fractal structure should the system reach equilibrium at large times.}

Let's collect all the potential terms, except the random force $f(t)$, acting on $\tilde{i}(r, t)$, that contribute to driving and damping infections in Eqn.\ref{rateofinfection} under a $V_{total}$ such that
 
\begin{equation}
V'_{total} = V'_{SEIR} + V'_{i} + V'_{D}
\label{totalV}
\end{equation}  

Control measures are switched on and off at certain times $t_{i}, t_{f}$ and can be imposed more than once. We account for their duration and time dependence by introducing a memory function in these potentials, a time window function $\theta(t -t_{i})$. An adiabatic (gradual) change of the restriction over time can be captured by a smooth hyperbolic tangent $Tanh(t_{i}, t)$ function, a sudden change by a Gaussian window function, and a change with a fixed start time but indefinite in the future is captured by a step function $\Theta(t - t_{i})$. By allowing the starting and ending time of these measures to be variable we can predict how the daily infection curves would change accordingly as a function of the time and duration of a control measure being imposed again the future. For example, a measure starting at $t_i$ but thereafter lasting indefinitely \footnote{By indefinitely we mean time scales from now to until we are immune to the virus}, would be described by a step function for their time window function.

\begin{equation}
V_{i} \longrightarrow V_{i}\Theta(t - t_{i})
\end{equation}

We allow the diffusion parameter $D = D(t)$ to vary with time in each region, rather than be a constant.
Moreover, a detailed analysis of the impact of border closings by region would require we introduce a function $g(\vec{r})$ which multiplies the noise term $f(t)$ so that $f(t) \rightarrow g(\vec{r}) f(t)$ and $\langle g(\vec{r})f(t) g(\vec{r'})f(t')\rangle = D(\vec{r}, t) \delta(t - t') \delta^{2}(\vec{r} - \vec{r'})$.

The stochastic differential Langevin equation, with all the time dependent parameters described above is the equation the machine learns to solve in predicting the infection curves as a function of time for each region. Coefficients $\tilde{\beta}$, $D(t)$ the diffusion parameter for random noise, $\tilde{\alpha}$ the lockdown confining potential, and $\delta$ the masks and social distancing parameter, are figured out by best fitting the solution with data provided daily by Johns Hopkins University. These learned parameters are then used to predict future daily infection curves for each region as a function of the restriction measures that may be introduced or relaxed in a region in the next twelve months. We illustrate the predicted daily infection curves given a particular choice of future measures ${(\tilde{\alpha}, \delta, \tilde{\beta}, \gamma)}$ as function of time for three US states below, and plan to post the predicted curves for any region in \cite{jmmmwebsite}. 

As a first step in using AI to implement this model, to gain in speed at the expense of complexity, we set the white noise term in the Langevin equation $f(t)$ to zero when solving Eqn.\ref{rateofinfection}. All the results shown in the next section and made available in \cite{jmmmwebsite} are based on the simplified Eqn.\ref{rateofinfection}. Without diffusion, the Langevin equation becomes a deterministic equation of 'motion' which takes far less time to solve numerically. We will report the results for the case $f(t) \ne 0$ in a sequel paper \cite{paperII}.

In the absence of the diffusion term at this stage, we account for the impact of closed borders by imposing absorbing boundary conditions, meaning no incoming or outgoing flux of people in a population at $\vec{r} = R_{country} = $ the border of states and countries. With the noise term taken to be zero in step one, to account for the part of diffusion of infection through a population that school openings may have contributed the machine absorbs its effect in the time dependence of the $V_{total}$ parameters $\tilde{\beta}, \tilde{\alpha}, \delta$. It should be noted that setting diffusion to zero converts the Langevin equation in a deterministic equation. As will be seen in the plots below, ignoring diffusion predicts the infection curves to drop to zero when the total force of infections $V'_{total} = 0$ in Eqn.\ref{totalV} and therefore ignores the probability that the danger of infection spread may persist even when restriction measures remain intact. 

Epidemiological studies and data have already shown the impact masks and distancing have on damping infection transmission. School kids below 18 years old make up $23\%$ of the population. To get an intuitive idea on how the machine in the absence of the diffusion term estimates the school effect and absorbs it as an increase in infection spread on the predicted curves, assuming that younger kids cannot effectively follow the guidelines or facilities are such that it is harder to social distance, then any increase \math{\tilde{\beta}} can be attributed to school openings. Currently we do not have a sufficient amount of data to isolate this effect in increasing the daily infection curves. However, given that all the other measures where adult population is concerned remain the same, then any increase in our predicted daily infection would be due to school openings. We plan to train the machine when more data on school openings is available and to use our model to quantify and report the increase in $\beta$ from school openings when all other restrictions are in place.  

\subsection{The Probability Distribution Function For The Complete Stochastic Model}

A more accurate description of the system's evolution $\tilde{i}(\vec{r}, t)$ is given by the complete stochastic mathematical model of a controlled pandemic of Eqn.\ref{rateofinfection} which includes the diffusion of infection through a random white noise term $f(t)$ for our restricted SAW random walker.  

For completeness let us provide the mathematical details of the model for the stochastic case when diffusion is included. We will show its AI implemented predictions in a future publication \cite{paperII}. We expect that with diffusion included the relative weights of the other parameters in $V_{total}$ as a function of time reported here may change. In contrast to the deterministic type Eqn. \ref{rateofinfection} of $D = 0$ solved in this work, in  the case of a diffusive system the solution is given by a probability distribution function (PDF) of infections $\tilde{i}(\vec{r}, t)$ is given by Eqn.\ref{fokkerplanck}, due to it stochastic nature. Let us denote the probability that our system $\tilde{i}(\vec{r}, t)$ which started at some initial position $\vec{r_{0}}, t_{0}$ is found at location $\vec{r}$ at some later time $t$ by $P[\tilde{i}(\vec{r}, t); \tilde{i}(\vec{r}_{0}, t_{0})] = P(\vec{r},t)$ for short, regardless of how it got there. The equation for the PDF derived from the above Langevin Eqn.\ref{rateofinfection} in $(2+1)D$ is known as the Fokker Planck equation.

The Fokker Planck equations for the probability distribution function $P(\vec{r}, t)$ of our self avoiding walker $\tilde{i}$ moving on a two dimensional space and restricted by the control measure contained in $V_{total}$ of Eqn.\ref{totalV} at any moment in time is given by (see \cite{fokkerplanck})

\begin{equation}
\dfrac{\partial P(\vec{r},t)}{\partial t} = \nabla_{\tilde{i}} [\nabla_{\tilde{i}}V_{total} P + D(t) \nabla_{\tilde{i}}P] = \nabla_{\tilde{i}} J.
\label{fokkerplanck}
\end{equation}

where again,

\begin{equation}
\nabla_{\tilde{i}} = \frac{d}{d\tilde{i}}
\end{equation}

and $V_{total}$ is given by Eqn.\ref{totalV} which enters the Langevin Eqn.\ref{rateofinfection}. The combination of the potential and diffusion terms on the right hand side of Eqn.\ref{fokkerplanck}, $J$, denotes the probability current of infection given by

\begin{equation}
 J = [\nabla_{\tilde{i}}V_{total} P + D(t) \nabla_{\tilde{i}}P]
\label{infectioncurrent}
\end{equation}

It should be noted that the probability of infection remains non zero when $D(t) \ne 0$ even when the total force of infection Eqn.\ref{totalV} in the Langevin equation is driven to zero from the impact of restrictions.

If equilibrium is achieved at large times, then $\nabla_{\tilde{i}} J = 0$ (see \cite{kardar,fokkerplanck,fokkerplanckneural}). It is only in this situation when the gradient of the infection current $\nabla J = 0$ that the risk of infection spreading goes.

The initial conditions for Eqn. \ref{fokkerplanck} are defined at $t=0$ when the first positive test is confirmed are denoted by $\tilde{i}(t = 0)$ = $i_{0}$, s(t = 0) = $s_{0}$. By training the machine on previous data the machine estimates the time derivative of the initial data ($\tilde{i_{0}}, \frac{d\tilde{i}(t=0}{d t}, s_{0}$) despite limited testing. Since the growth is exponential early on than the time derivative of $\tilde{i}$ at initial time $t=0$ has a simple relation to $\tilde{i}$ given by the slope of that curve.

Note that for $D(t)=0$ if $[\tilde{\beta} - \tilde{\alpha} - \delta] \le 0$ in the Langevin equation, Eqn.\ref{rateofinfection}, than the control measures introduce in $V_{total}$ of Eqn.\ref{totalV} effectively shut down the spread of infections (see plots in section 4). However this is not strictly the case in a stochastic model where diffusion introduces a nonzero probability that some amount of re$-$infection can enter a region through travel or may still be present and diffuse through a population. The reason for this behavior is the fact that the random diffusion captures the unknown and unreported elements in the population which somehow escaped the healthcare net, who should have been quarantined but weren't, such as some people who may be asymptomatic and careless in following the measures $V_{total}$ introduces, or those who cannot afford to go to the doctor or miss work, as well as any accidental exposure to infection through touching an infected surface or airborne particles. Since diffusion captures all the random elements that cannot be anticipated in $V_{total}$, the stochastic model gives a more realistic description of the spread of infection, than the deterministic model.  

No analytical solution exists for this class of stochastic models of a restricted self avoiding random walker in two spatial dimensions. Due to their complexity these models are also notoriously hard to solve numerically within a short time. Therefore, during an unfolding pandemic when time is of essence, AI implemented results of Eqn.\ref{fokkerplanck} which we will show in the sequel paper \cite{paperII} offer a powerful way for obtaining numerical solutions for the PDF of infection and for the diffusion parameter $D(t)$ within a reasonable time,  at any point $(r, t)$ in spacetime. 

It is well known that, for the case when the potential parameters $\tilde{\beta}, \tilde{\alpha}, \delta$ are constants, these stochastic models lead to critical phenomena \cite{kardar,kardar2} at $t ->\infty$. In this case, for $\nabla J=0$ the infection spread $\tilde{i}(r,t)$ would follow a fractal structure of self similar 'hot'and 'cold' clusters of infections of dimension less than two. In the limit of large times $t \rightarrow \infty$, the solutions to Eqn.\ref{fokkerplanck} have a universal scaling behavior and for large populations, $N \rightarrow \infty$, they approach a scale free network. In two dimensions, critical exponents are known numerically but not rigorously derived analytically. Furthermore, in equilibrium $\nabla J = 0$ the diffusion parameter $D$ of infection and the dissipation of infection $\mu$ are related by a form of the fluctuation dissipation theorem (the Einstein-Schomlowsky relation) as
$\frac{D}{\mu} \approx \textit{constant}$.

A way to understand the above relation between diffusion and dissipation of infection is to recall the relation between the rate of change of the infected field $\tilde{i}$ (our system) and the susceptible parts of the population (the 'background') $\frac{d \tilde{i}}{d t} = - \frac{d s}{dt}$ of Eqns.(\ref{infectioneqn}, \ref{rateofhealthy}, \ref{rateofinfection}).

The scaling behavior at large times for the 'mean square distance' (MSD) for these systems in two dimensions in terms of the critical exponents expected to go as

\begin{equation}
\langle \tilde{i}^{2} \rangle = (4Dt)^{2\nu} = (4Dt)^{3/2}
\end{equation}

\begin{equation}
\langle \tilde{i} \rangle = \sqrt{\langle \tilde{i}^{2} }\rangle \sim \langle \tilde{i} \rangle_{0} t^{3/4} 
\end{equation}

Note that for a diffusive stochastic system, the infection growth at late times is polynomial with time rather than exponential growth like the in the $SEIR$ model.
In this section we reported the known characteristic features and scaling behavior in literature \cite{2Dsaw} for these stochastic models as $t \rightarrow \infty$ when the system reaches equilibrium $\nabla J=0$. However we should caution that the results for the large time evolution of the system and the values of critical exponents studied in literature, do not have time dependent parameters in $V_{total}$ as we have here. The time dependence of the coupling constants, in addition to diffusion, adds an extra layer of complexity to the stochastic system. Furthermore, non-stationary solutions $\nabla J \ne 0$ for $P[\tilde{i}(r, t)]$ for a SAW model with time dependent parameters are extremely hard and can only be solved numerically by powerful machines. 
This is a highly complex system which will require considerably more time and computing power. We are currently preparing to train the machine for the full stochastic model with the time dependent coefficients presented here. We will report the solutions of Eqn.\ref{fokkerplanck} for the probability distribution function of infections and estimate the diffusion parameter $D(t)$ as a function of time in a future publication \cite{paperII}.

\section{Methods}

The model in \r{rateofinfection} has several parameters which are unknown:
 $\tilde{\beta}, \tilde{\alpha}, \delta$, for ease of notation from here we will take  $\tilde{\beta}, \tilde{\alpha}, \delta$  to be  \math{\beta, \alpha, \delta} and drop the tilde, but we emphasize that future references of \math{\beta, \alpha, \delta}do absorb and reflect the region and time dependent version of the parameters. To use the model for forecasting and
scenario analysis requires us to identify these parameters.
To identify these parameters, we fit the model to observed daily confirmed
infections obtained from John's Hopkins University, \cite{JHU-Data}.
The parameters may vary from one region to another, hence we train the machine and best fit the parameters of the model independently to each region's data.
To fit the model to data, we make two practical modifications:
\begin{inparaenum}[(i)]
\item
  Infections data is reported daily, so we convert the differential
  equations to finite difference equations.
\item
  Data only reports confirmed (usually symptomatic)
  infections. An infection is only
  confirmed after a time-lag \math{k}-days, and further, only a fraction
  \math{\gamma} of
  the infections are confirmed. The parameter \math{k} corresponds approximately to the incubation time of the virus and we set it to 8 \cite{IncubationTime}. The parameter
  \math{\gamma} is important in linking the observed infections to the
  resevoir of all infections.
\end{inparaenum}
To expedite the fitting process, in this paper we only consider the diffusionless $D(t) = 0$ simplified version of the model presented in the above section.
Taking \math{D=0} removes the \math{f(t)} term in \r{rateofinfection}.
We also set \math{\zeta=2} and take \math{r\approx i^{-1/2}} (see the
discussion in section 2). Let us first discuss the
finite difference model with a lag.

{\it Finite Difference Model:}

It is convenient
to introduce a new state
variable \math{q(t)} to denote the number
of {\it newly} created infections at time \math{t} in region \math{r}.
\math{q(t)} approximately corresponds to the rate
\math{d\tilde i/dt} of Eqn. \r{rateofinfection} in region \math{r}.
Recall that \math{\tilde i(t)} is the cumulative
total number of individuals up to time \math{t} in 
who have been infected by the
disease (including those who have
recovered), and \math{e(t)}
is the number of exposed and currently infectious individuals at time
\math{t}. At time \math{t}, let \math{s(t)} be the number of
susceptible individuals. We introduce two additional
variables \math{r_s(t)} and \math{x(t)}, which have no exact counterpart in the
standard SEIR model. \math{r_s(t)} is the cumulative
number of individuals up to time
\math{t} who contracted the disease at some earlier time and have now
{\it silently} recovered. Silently means without detection, usually
because there were no symptoms.
\math{x(t)} is the cumulative number of {\it confirmed} infections up to
time \math{t}. The contribution to
\math{x(t)} at time \math{t} comes from the fraction of new infections at time
\math{t-k} which get confirmed at time \math{t}
(the remaining fraction silently recover). Although we did not explicitly show the \math{\vec{r}} dependence in the variables defined above it should be understood that they are functions of space as well as time. The coupled
finite difference equations are:
\eqar{
  q(t)&=& \beta e(t-1)s(t-1) - \alpha e(t-1)-\delta e(t-1)(1-2e(t-1))
  \label{eq:diff1}\\
s(t)&=& s(t-1)-q(t)\label{eq:diff2}\\
e(t)&=& e(t-1)+q(t)-q(t-k)\label{eq:diff3}\\
x(t)&=& x(t-1)+\gamma q(t-k)\label{eq:diff4}\\
r_s(t)&=& r_s(t-1)+(1-\gamma)q(t-k)\label{eq:diff5}
}
Equation \r{eq:diff1} contains the infectious force and damping
potentials from \r{rateofinfection} which control
the rate at which new infections are produced,
\math{q(t)}.
Equation \r{eq:diff2} merely says that the susceptible
population decreases by the new infections.
Equation \r{eq:diff3} says that the infectious individuals 
increase by the new infections minus the infectious individuals who
recovered silently or who
became symptomatic and got discovered (and quarantined).
Due to the lag \math{k} the loss of infectious individuals is exactly
\math{q(t-k)}.
Equations \r{eq:diff4} and \r{eq:diff5} say that a fraction
\math{\gamma} of the infectious get symptomatic and discovered and
the other \math{(1-\gamma)}-fraction silently recover without serious symptoms.
Recall that \math{\tilde i(t)=e(t)+x(t)+r_s(t)}.
It is only \math{x(t)} which is observed.

The parameters \math{(\beta,\alpha,\delta,\gamma)} are unknown.
From data provided by John's Hopkins University \cite{JHU-Data},
we observe very noisy estimates \math{\{\hat x(0),\hat x(1),\hat x(2),\ldots, \hat x(T)\}},
where timestep 0 indicates the first {\it confirmed} infection.
We set \math{i(-k)=x(0)/N} where \math{N} is the population of the
region being analysed, and correspondingly
\math{s(-k)=1-i(-k), q(-k)=i(-k), x(-k)=0, r_s(-k)=0}.
Given these initial conditions, the trajectory of \math{x(t)} is determined
given \math{(\beta,\alpha,\delta,\gamma)} and
Equations \r{eq:diff1}--\r{eq:diff5}. Hence we
identify \math{(\beta^*,\alpha^*,\delta^*,\gamma^*)} as the parameters
which minimize the mean squared error between the observed trajectory
\math{\bm{\hat x}} and the model's trajectory
\math{\bm{x}(\beta,\alpha,\delta,\gamma)}. That is,
\mldc{
  (\beta^*,\alpha^*,\delta^*,\gamma^*)
  =
  \argmin_{(\beta,\alpha,\delta,\gamma)}\sum_{t=0}^T(x(t|\beta,\alpha,\delta,\gamma)-\hat x(t))^2.
\label{eq:MSE}
}

{\it Lockdown, Social Distancing and Masks:}

Again, to address the practical situation with COVID-19, we
have to address the effects of lockdown, social distancing and masks.
Indeed, this is one of the advantages of our model, that it
gives us simple switches for turning on and off
various social distancing measures.
We assume that lockdown begins at some time
\math{\tau} and lasts till \math{\tau+L}. In general fitting of the data,
the lockdown period
\math{L} is about 90 days for all USA regions. Naturally, this parameter
\math{L}
is tuned to each specific region and it is calculated by the machine learning algorithm from JHU data \cite{JHU-Data}. The start of the
lockdown, \math{\tau}, is determined by a robust
changepoint (A sample or time instant at which some statistical property of a signal changes abruptly) analysis in the
time-series (for example see \cite{MalikPaper}). 

Parameters representing the strength of each measure on any region vary with time and duration. Being trained on past data, the machine learns the time and strength of these measures $\alpha, \delta$ and the remaining degree of randomness $\beta$ and the time when they were introduced suddenly in the past. Therefore, we include that time dependence in 
\math{\beta, \alpha, \delta} in \r{eq:diff1} by breaking down their time dependence in three time intervals, namely: the 'random spread time',i.e. the time before the lockdown $t < \tau$ when all restrictions are zero and the virus is spreading randomly, i.e. when $\beta = \beta_{max}$; the 'lockdown duration', i.e. the time interval during the first full lockdown $\alpha = \alpha_{max}$ being in effect $\tau < t < \tau + L$ in a region. \math{L} should be understood to be a function of space and time \math{L(\vec{r},t} since the lockdown time and duration varies by region, and the machine can learn to read this parameter from data. Social distancing and masks restriction $\delta =\delta_{max}$ are also in full effect during this time and can extend beyond. As a result, $\beta$ drops from its maximum value by a factor $\lambda_{m}^{p} \beta$ where the reduction parameter $\lambda_{m}^{p}$ in this interval is estimated from past data from the machine; and, future time interval which can extend for two years, i.e. $t > \tau +L$ when some of the restrictions are relaxed or not followed rigorously. For example, although lockdown officially ended in May, the machine estimates that its impact parameter $\alpha$ on infection spread was still at half its strength until August because many people were encouraged and continued to work remotely, schools were closed during summer, and travel and hospitality industries were at a minimum, therefore many people stayed home. The future increase or decrease in $\beta$ is captured by the parameter $\lambda_m$ and it depends on future decisions taken on restrictions in any region. The strength of social distancing follows similar patterns in past data and it is learned by the machine. The change, for example decrease, in the strength of lockdown and social distancing restrictions that can be imposed in the future on any region is captured by the region and time dependent parameters $\lambda_{L}(t), \lambda_{sd}(t)$ respectively. The time dependence on these parameters allows us the freedom to predict what happens to the daily infection curves if restrictions are introduced at certain times in the future. All past data parameters are best fitted by the machine. With these definitions, we have    
\eqar{
  \beta(t)&=&
  \begin{cases}
    \beta_{max}&t<\tau;\\
    \lambda_{m}^{p}\beta&\tau\le t\le\tau+L;\\
    \lambda_m \beta&t>\tau+L.\\
  \end{cases}
  \label{eq:lock1}
  \\
  \alpha(t)&=&
  \begin{cases}
    0&t<\tau;\\
    \alpha(t) &\tau\le t\le\tau+L;\\
    \lambda_{L}(t)\alpha&t>\tau+L.\\
  \end{cases}
  \label{eq:lock2}
  \\
  \delta(t)&=&
  \begin{cases}
    0&t<\tau;\\
    \delta(t) &\tau\le t\le\tau+L;\\
    \lambda_{sd}(t)\delta&t>\tau+L.\\
  \end{cases}
  \label{eq:lock3}
}
Equation \r{eq:lock1} captures the effects of control measures of \math{\alpha} and \math{\delta} in reducing \math{\beta} after lockdown and in the future. Although \math{\beta} does not return to its maximum initial value in the future, it increases by a percentage defined by the factor \math{\lambda_m} when restrictions ease.  Equations \r{eq:lock2} and \r{eq:lock3} capture changes over time in the impact of lockdown and social distancing
potentials on damping infection spread. 

The step functions in Equations \r{eq:lock1}--\r{eq:lock3}
can be smoothened to better approximate reality, since control measures
are not instantly adopted throughout the whole population (except perhaps
a government mandated lockdown or curfew).
A transition from say a function \math{a(t)} to another function \math{b(t)} at time \math{t_0} can be
compactly written as a single function
\math{a(t)+\Theta(t-t_0)(b(t)-a(t))}, where \math{\Theta(\cdot)} is the
Heaviside threshold function. One can smoothen
this by replacing
\math{\Theta(t-t_0)} with any smooth approximation, such as
\math{\Theta(t-t_0)\approx (1+\tanh(c(t-t_0)))/2} for \math{c>0}
(smaller \math{c} is smoother) and the machine can be set to determine the value of c.

In summary, the lockdown start-time \math{\tau} is determined by the first
changepoint in a robust changepoint analysis of the time series
of confirmed infections.
For example, in the US the lockdown duration is about \math{L=90} days, but the machine learns this duration from data for any region. 
We fit the
parameters \math{\beta(\vec{r}, t),\alpha(\vec{r}, t),\delta,\gamma(\vec{r}, t),\lambda_{i} (\vec{r}, t)} (where $i$ stands for $m, L, sd$ ) to the data by
minimizing the MSE in \r{eq:MSE} using an exhaustive search over the ranges
\math{\beta\in[0.1,1]},
\math{\alpha\in[0,0.3]}, \math{\delta\in[0,0.2]},
\math{\gamma\in[0,0.3]}
and \math{\lambda_m\in[0.5,0.7]}.

By changing future strength of restrictions given \math{\alpha(t)}, \math{\delta(t)} and \math{\beta}, allows us to predicts and investigate
future scenarios in which the different social distancing measures are
relaxed at certain times. We will demonstrate some of these scenarios next in the Results section.

\section{Results}

In addition to the new mathematical model we present in this paper, we have made the code publicly available in \cite{ourcode}, including the machine learning algorithm. Data file is available daily from \cite{JHU-Data}. The machine is trained in every region of the world, therefore our model implemented by the deposited code can be used for predicting daily infection curves for any region in the world. We plan to post infection curves
for every region in the world on our website~\cite{jmmmwebsite}, for the two future control measure scenarios
described below. Here, we discuss in detail the results for
three states:{\it North Carolina, New York and Florida}.

These results are obtained by fitting the model to the region's
  reported data in \cite{JHU-Data}, shown by the red dots in the figures below, to obtain
  the coupling parameters \math{\beta(t),\alpha(t),\delta(t),\gamma,\lambda_m}. We show predicted curves in these three states by considering these two hypothetical future scenarios:

\renewcommand{\labelenumi}{\roman{enumi}}
  \begin{enumerate}[label=(\alph*),nolistsep]
  \item {\it Scenario A.}
    The official lockdown of last Spring ends. However population remains in a semi lockdown state with a decrease of 50\% in the impact of this restriction in slowing down infection spread, that is
    \math{\alpha\rightarrow\alpha/2}. Social distancing is extended for a short time interval (approximately from the end of lockdown until August 2020) after official lockdown ends, before this restriction too decreases to about 50\% . The exact time intervals when these measures are reduced can be seen in the plots below. The reduction in these restrictions occurs around the anticipated start of K-12 schooling in each state, assuming that social distancing measures will  consequently decrease and \math{\beta, \lambda_{m}} will increase due to the difficulty enforcing these measures in schools. 
  \item {\it Scenario B.}
    The same as above but social distances is maintained for a longer extension of time (extension past August 2020 into January 2021) before being reduced to 50\% impact. In this scenario the reduction in social distancing and therefore increase in  \math{\beta} starts happening in January 2021. This is done under the assumption that states can maintain restrictions such as social distancing and masks without a strict lockdown through January, for example by moving K-12 schools to a remote learning model for the Fall Semester. This assumption does not include the possibility of universities operating in person classes, if we were to assume that college students would follow guidelines like the rest of the adult population.
  \end{enumerate}

We are not proposing that these scenarios be adopted by policy makers, we are simply using these two examples to illustrate the predictive power of our mathematical model and its machine learning implementation.

One could instead
analyze any combination of future restrictions such as additional lockdowns,
partial mask compliance, etc.

Note that a different choice and/or duration of future restrictions
in a region will obviously produce very different daily infection curves. The danger of infection spread can be postponed but will continue to loom over the   susceptible population until a vaccine is developed. 
Hence, it is up to health-economists to evaluate the
human and economic loss of different scenarios to come up with the
optimal redistribution of the infection counts to tradeoff healthcare-system
overload against prolonged economic slump.
 Our Ai implemented predictive model can help in this effort by
efficiently determining months in advance the infection spread dynamics as a function of
future decisions on the schedule and strength of control measures.

We illustrate these predictions next.

\subsection{North Carolina, New York and  Florida}

\begin{figure}[H]
  \begin{tabular}{cc}
    \begin{tabular}{l}
        
      \includegraphics[width=0.475\textwidth]{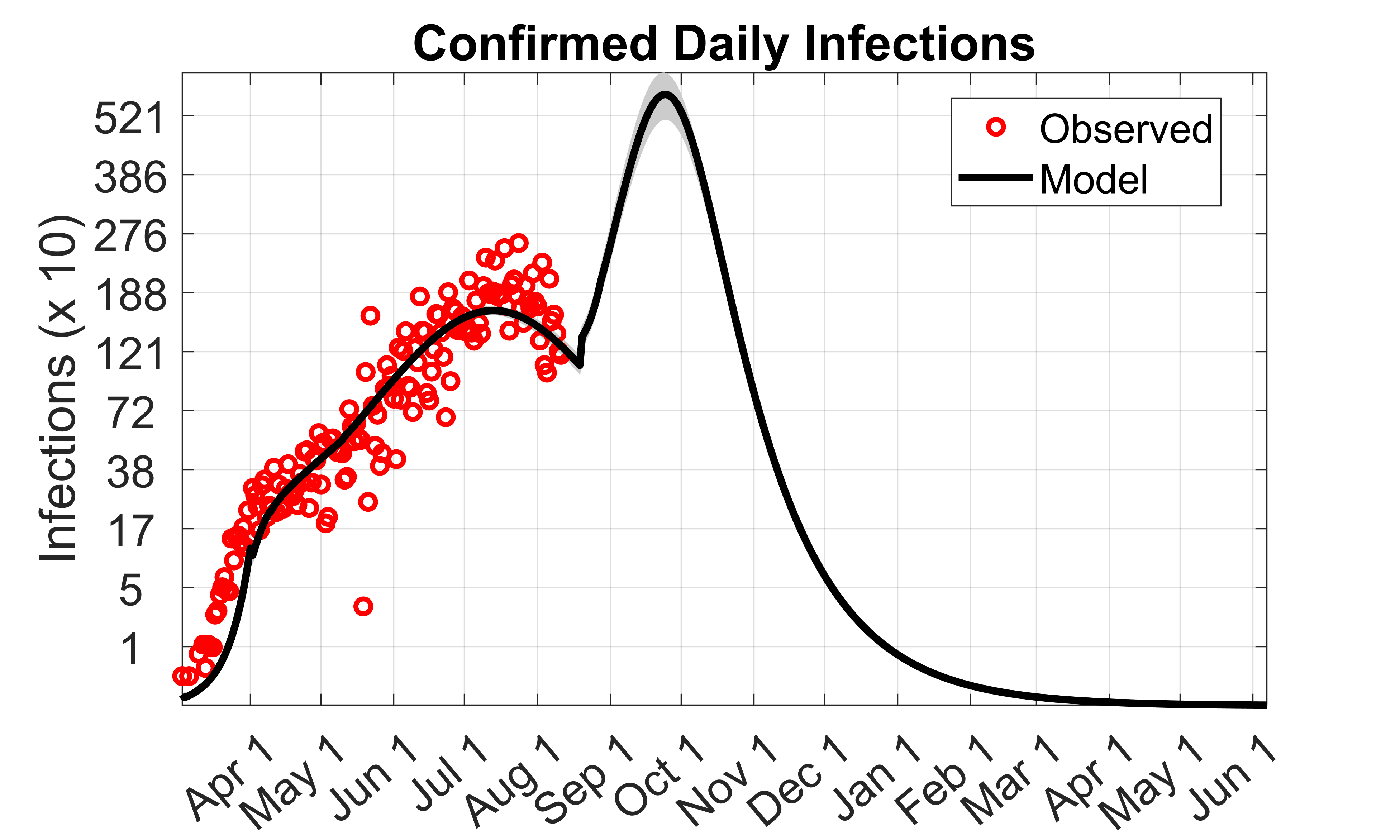}\\
      \includegraphics[width=0.475\textwidth]{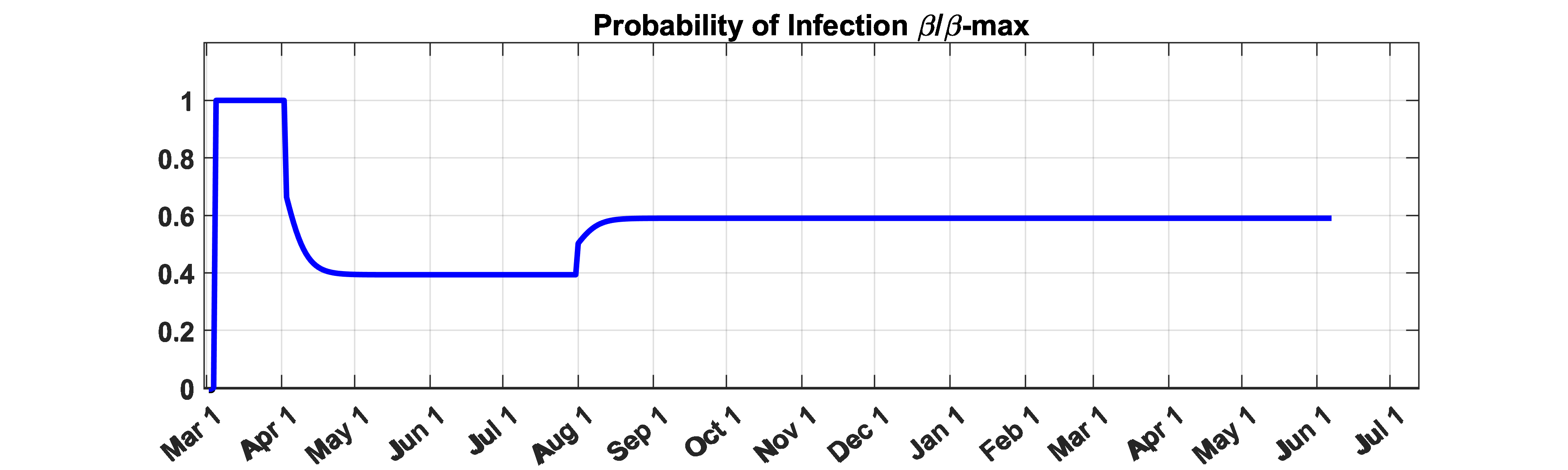}\\
      \includegraphics[width=0.475\textwidth]{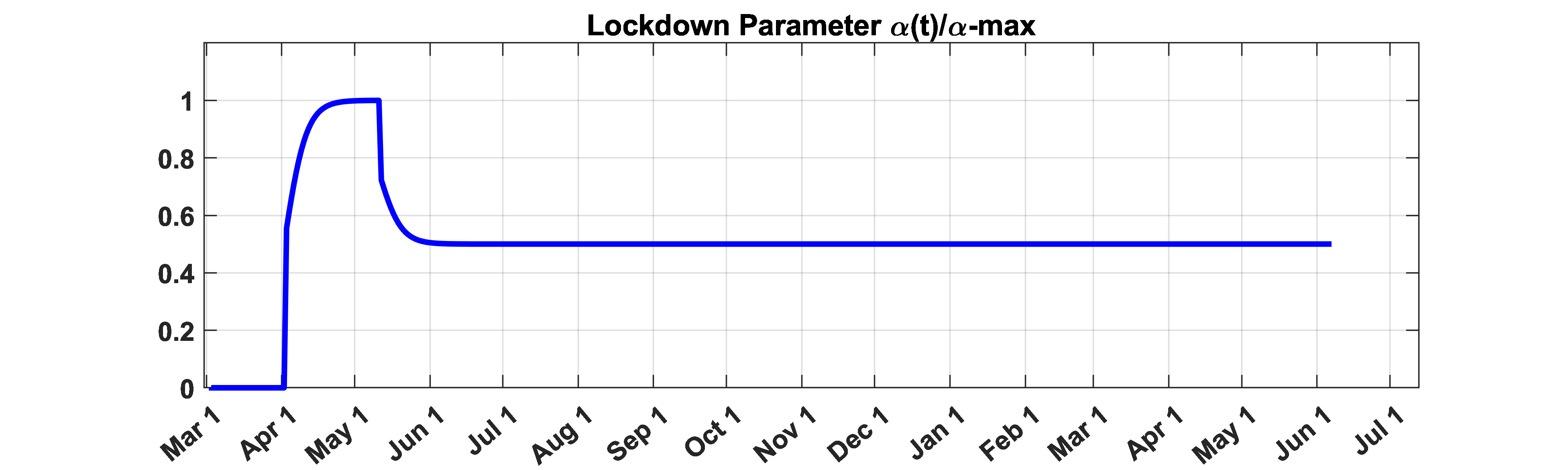}\\
      \includegraphics[width=0.475\textwidth]{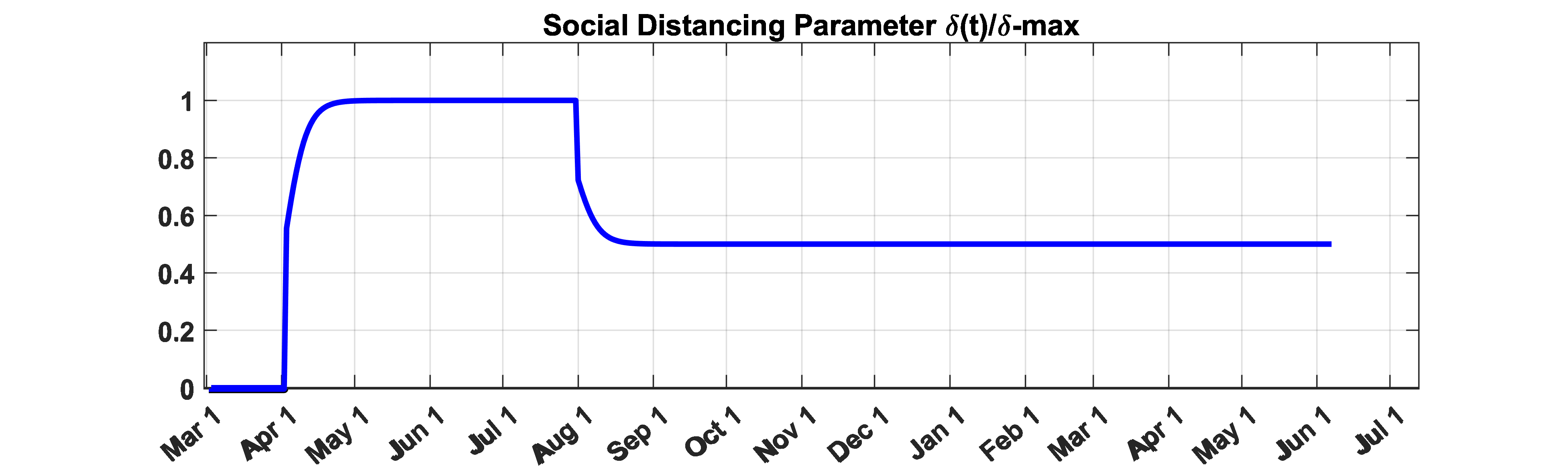}\\
      \includegraphics[width=0.475\textwidth]{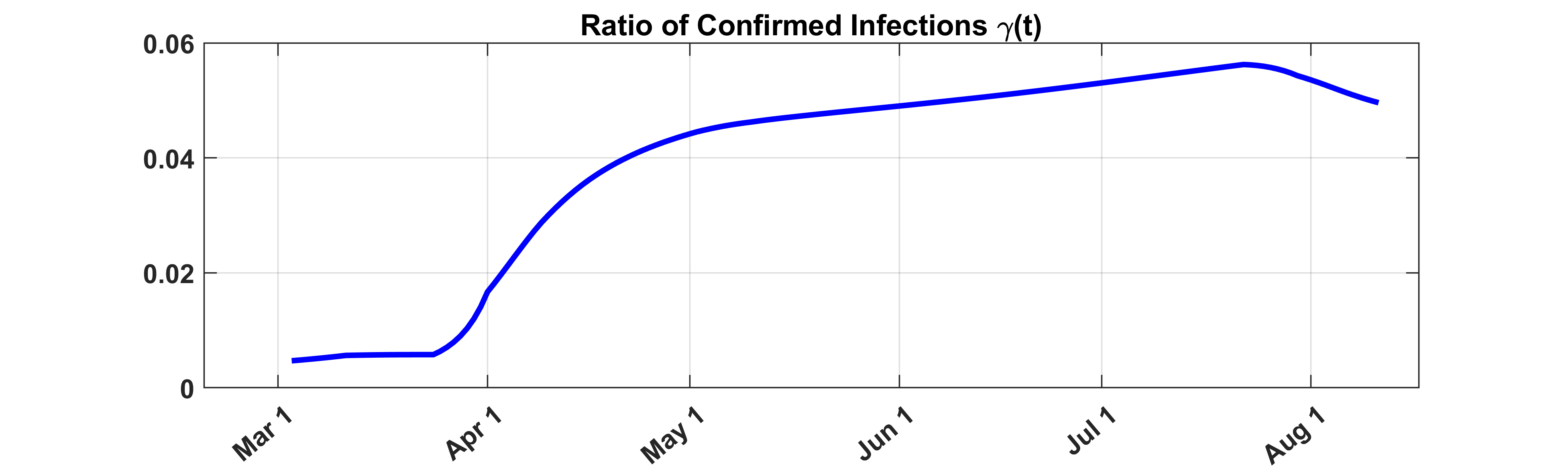}
    \end{tabular}
    &
    \begin{tabular}{l}
      \includegraphics[width=0.475\textwidth]{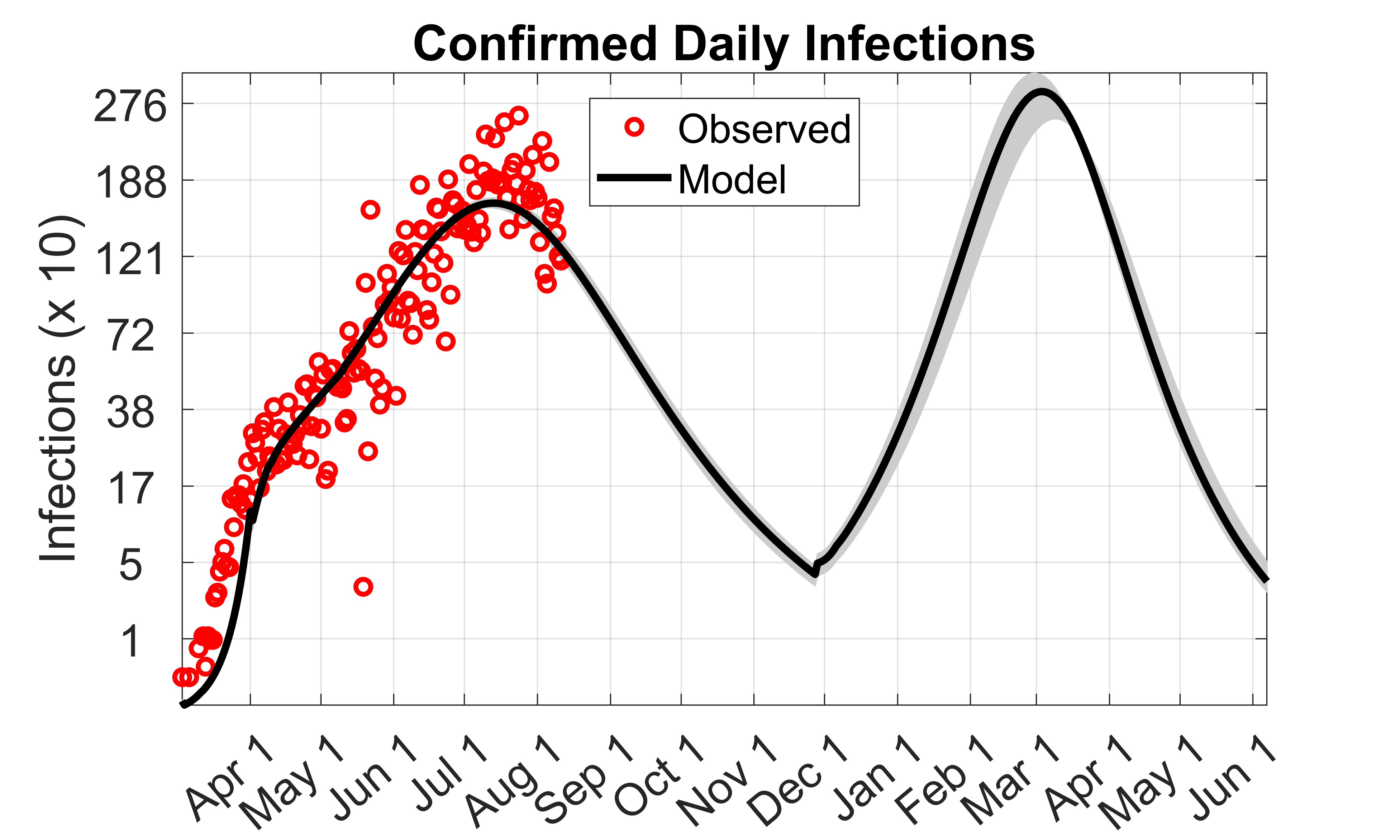}\\
      \includegraphics[width=0.475\textwidth]{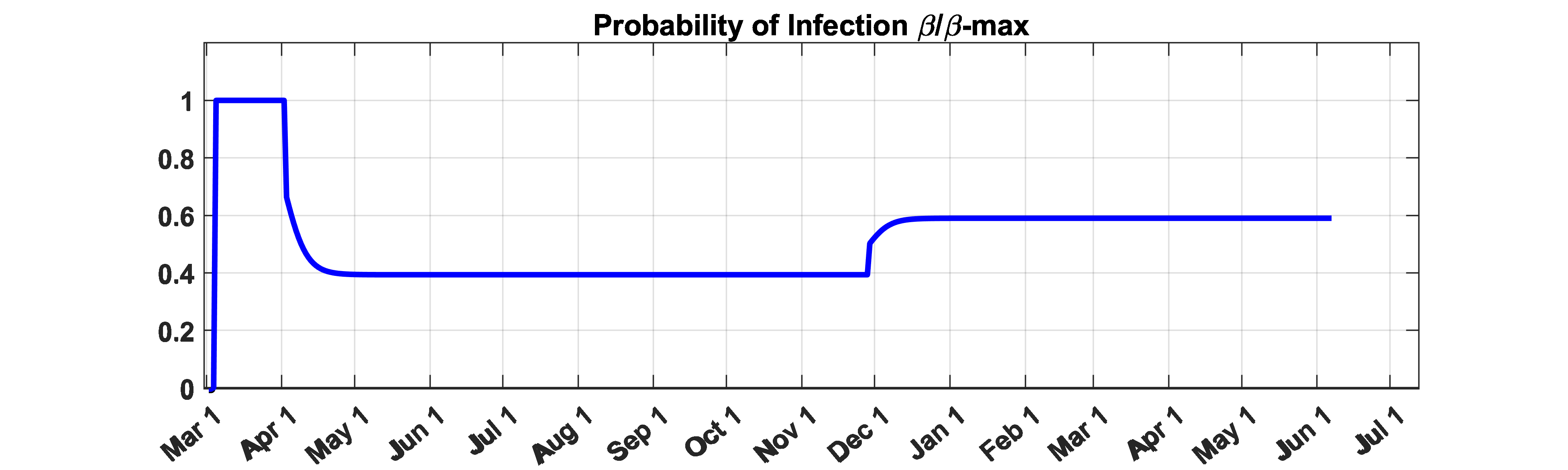}\\
      \includegraphics[width=0.475\textwidth]{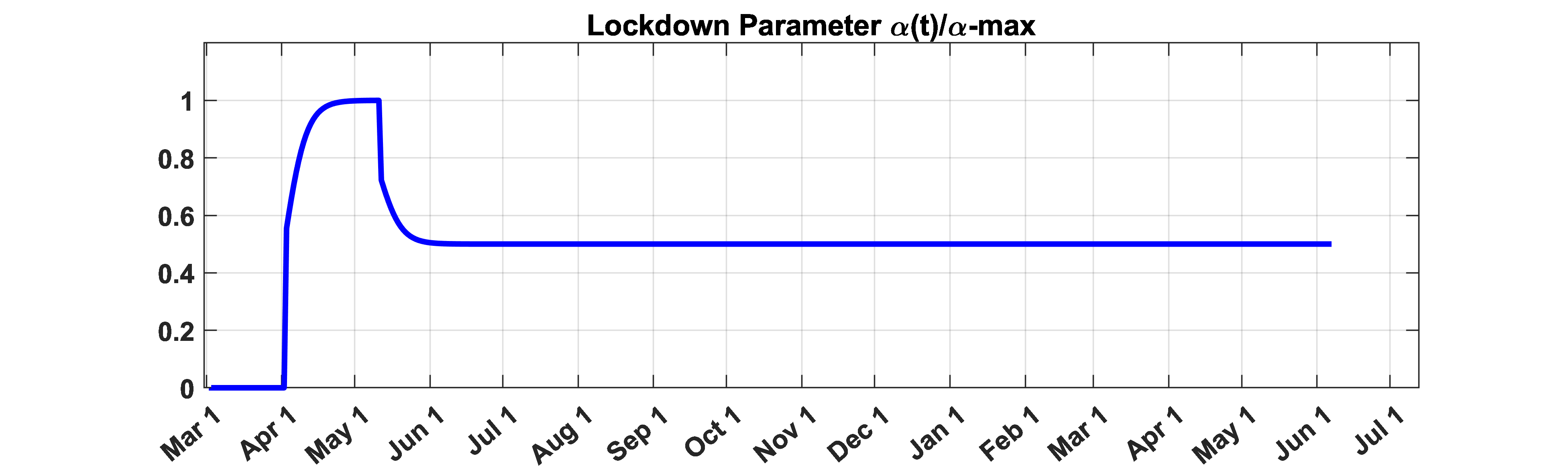}\\
      \includegraphics[width=0.475\textwidth]{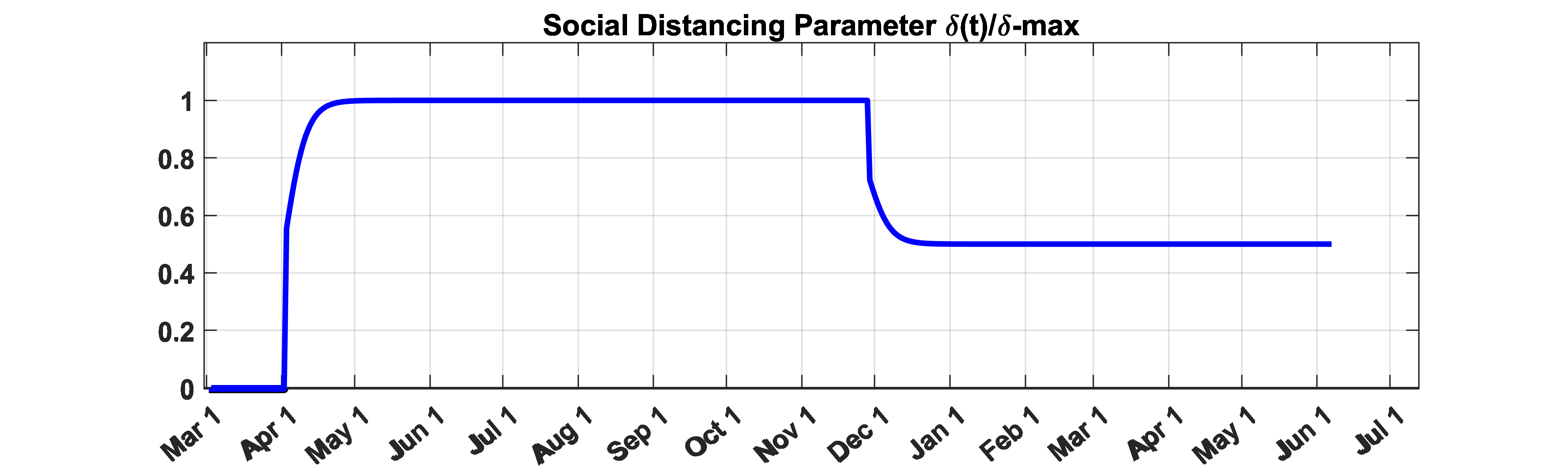}\\
      \includegraphics[width=0.475\textwidth]{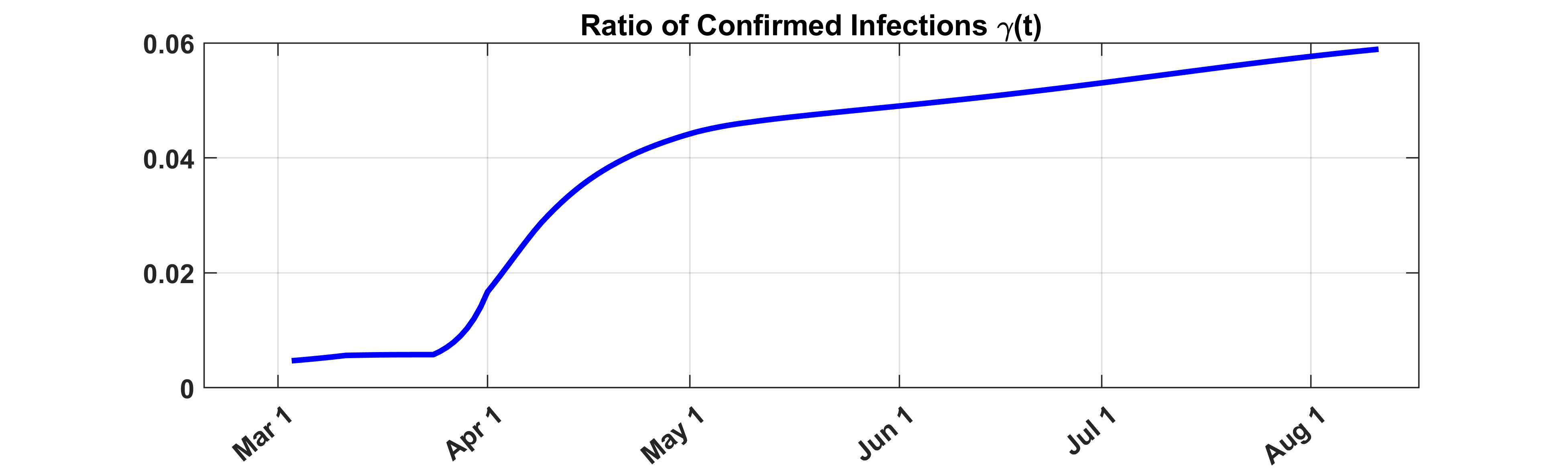}
    \end{tabular}\\
    (a) Scenario A.
    \\
    &
    (b) Scenario B. 
    \\
  \end{tabular}
  \caption{{\bf North Carolina (NC).}
    (Left) Scenario A relaxes control measures in accordance with real time policy decisions from each state \cite{JHU-Data,atlantic}. This results in relaxation of the lockdown measures on May 8, with social distancing and \math{\lambda_m} being relaxed around 80 days later, resulting in the spike of infections seen around October. 
  (Right) Scenario B extends social distancing and \math{\lambda_m} an additional 120 days, accounting for a scenario in which K-12 schools move to a remote learning model (as is the case in NC) in the fall 2020, and so the spike in infections is delayed until March 2021 should measures be relaxed next Spring.
Prolonging restrictions reduces \protect\ensuremath{\beta} and daily infection curves. The multiplying parameter for the reservoir \protect\ensuremath{\gamma} is also shown below.In both scenarios \protect\ensuremath{R_{0} = 3.07, \alpha_{max} = 0.01, \delta_{max} = 1.05e -08, \gamma_{today} = 0.063}, which corresponds to roughly 1 in 15 infections being confirmed.}
  \label{NCplots}
\end{figure}

\begin{figure}[H]
  \begin{tabular}{cc}
    \begin{tabular}{l}
      \includegraphics[width=0.475\textwidth]{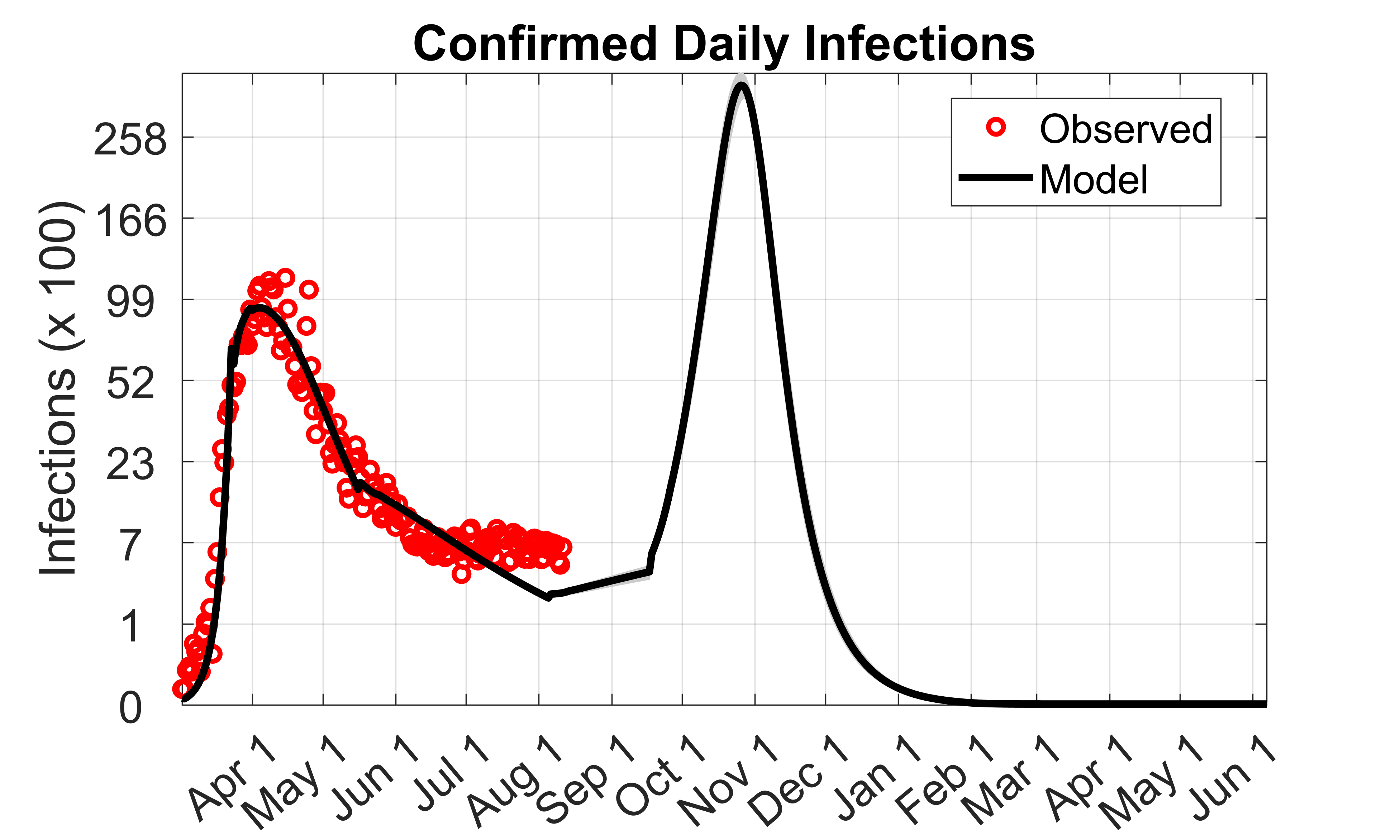}\\
      \includegraphics[width=0.475\textwidth]{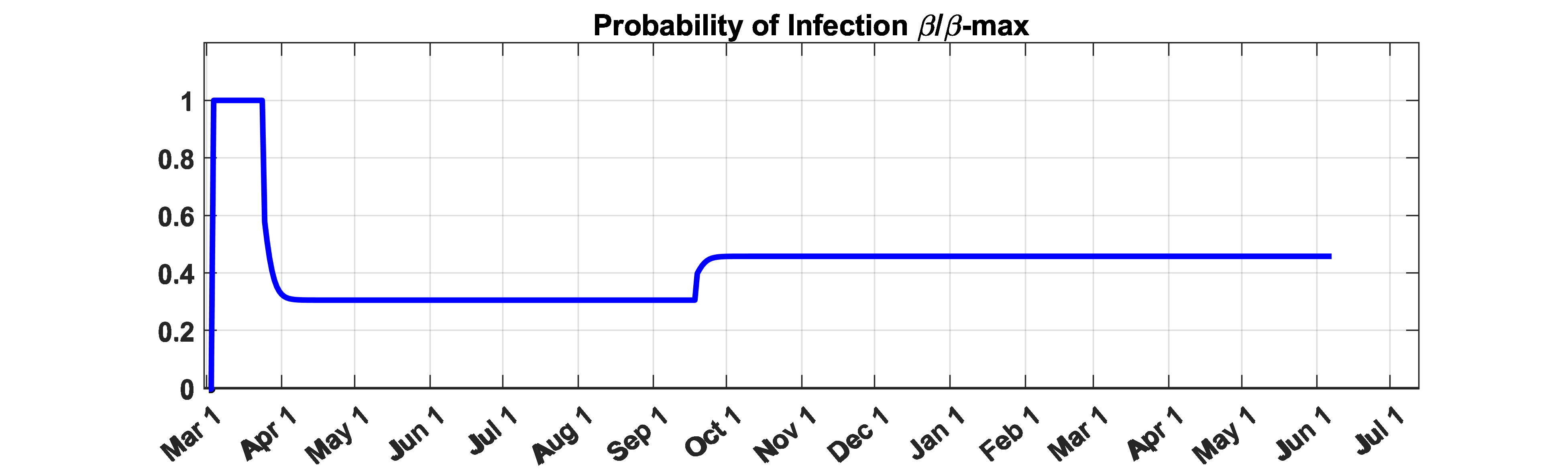}\\
      \includegraphics[width=0.475\textwidth]{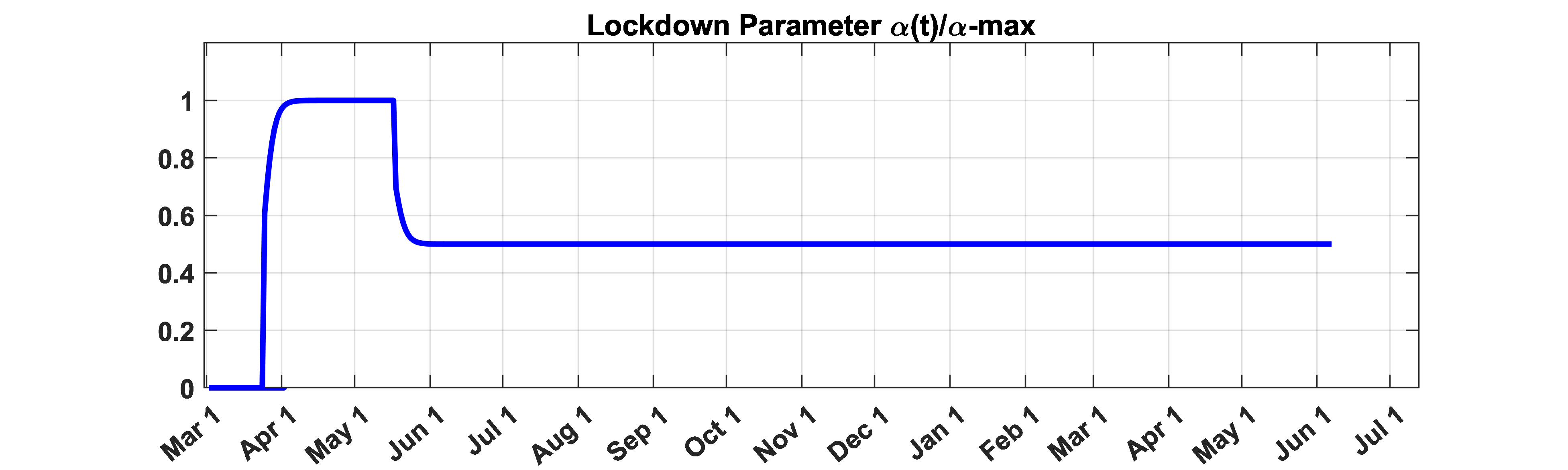}\\
      \includegraphics[width=0.475\textwidth]{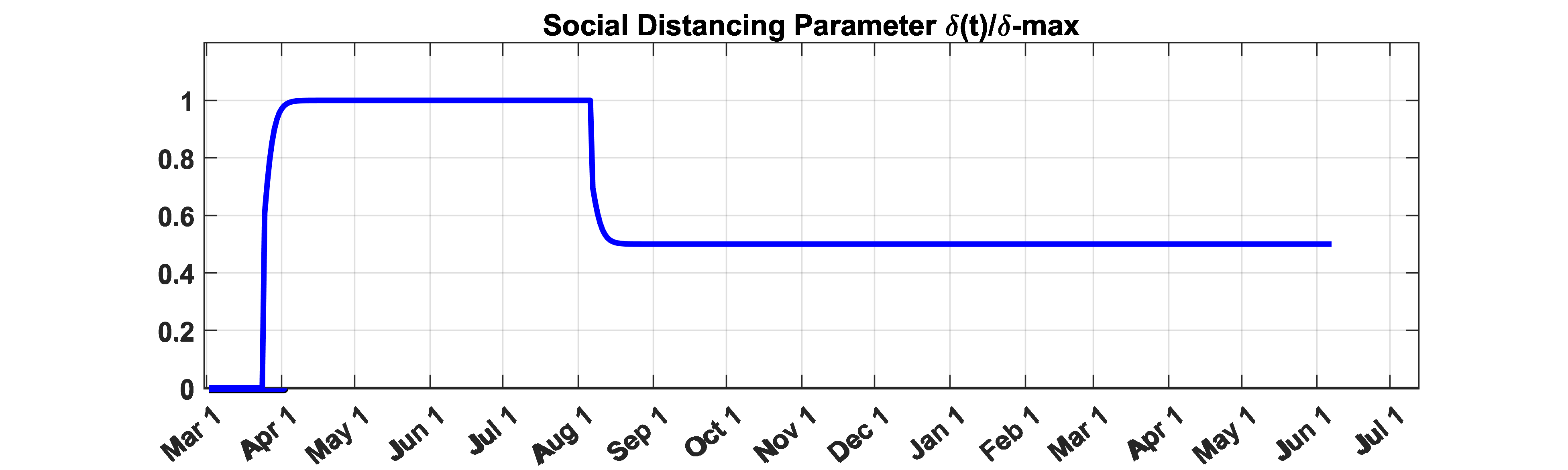}\\
      \includegraphics[width=0.475\textwidth]{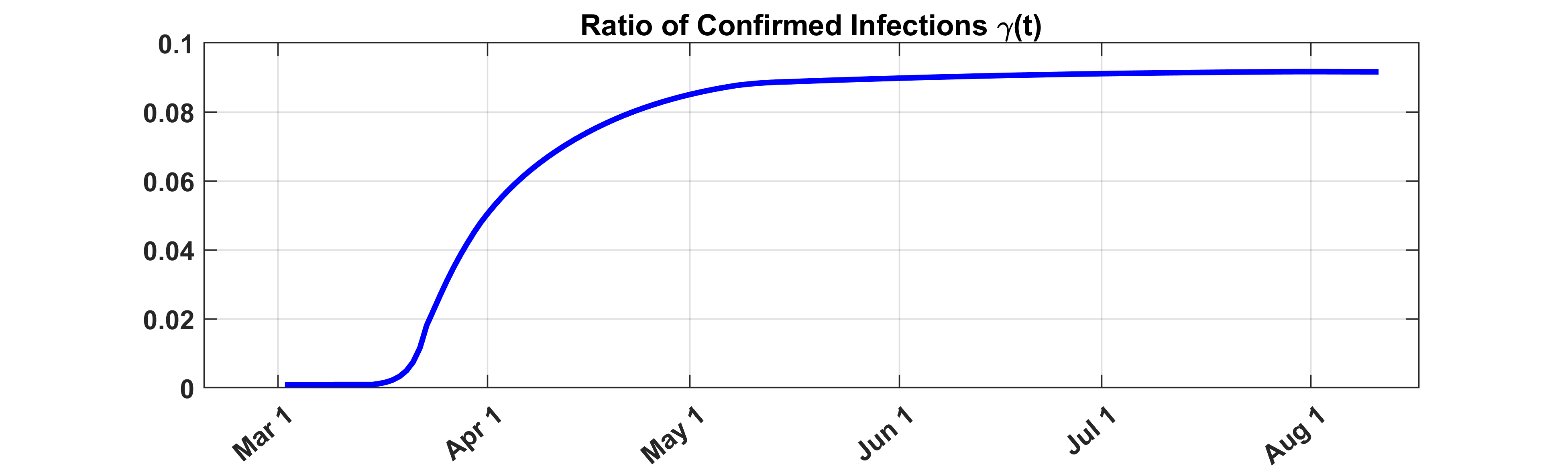}
    \end{tabular}
    &
    \begin{tabular}{l}
      \includegraphics[width=0.475\textwidth]{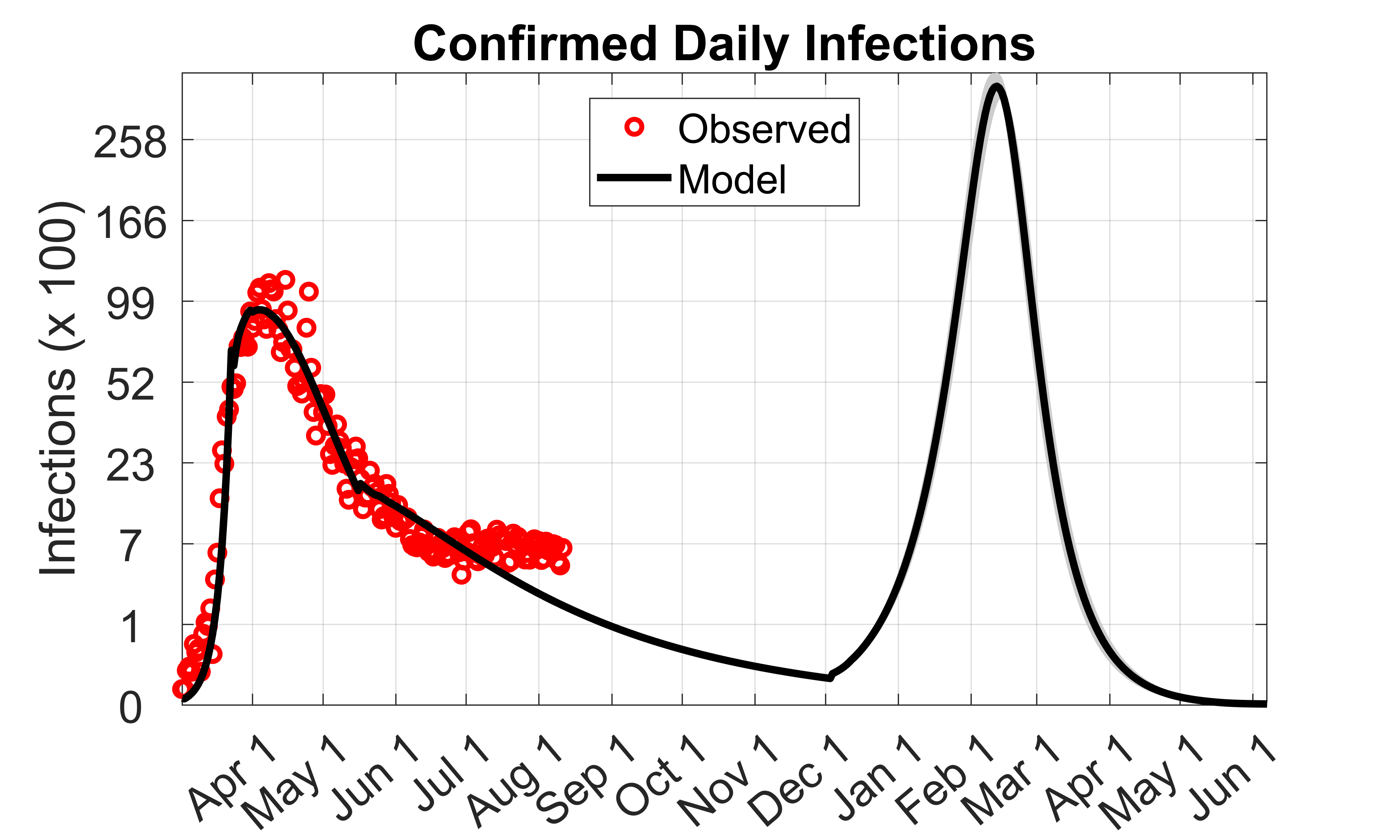}\\
      \includegraphics[width=0.475\textwidth]{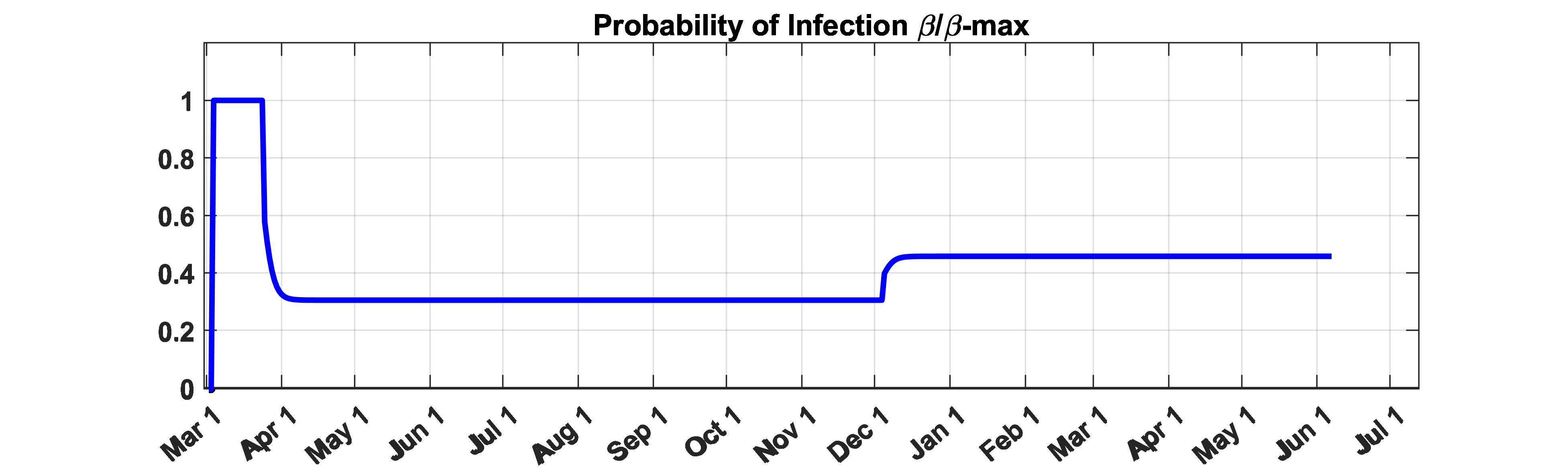}\\
      \includegraphics[width=0.475\textwidth]{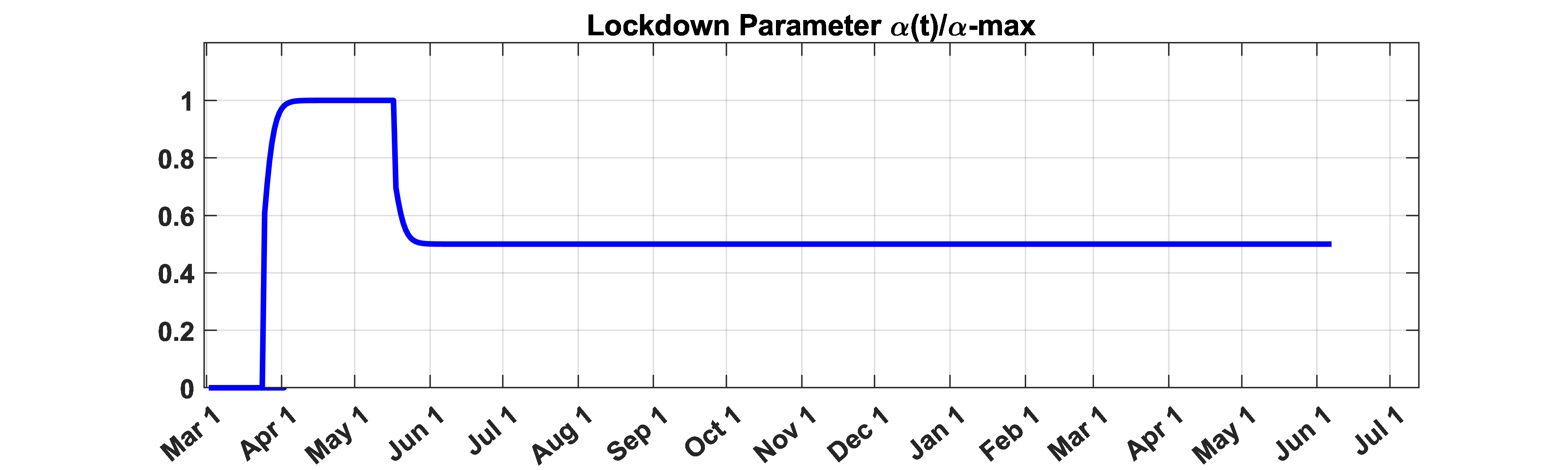}\\
      \includegraphics[width=0.475\textwidth]{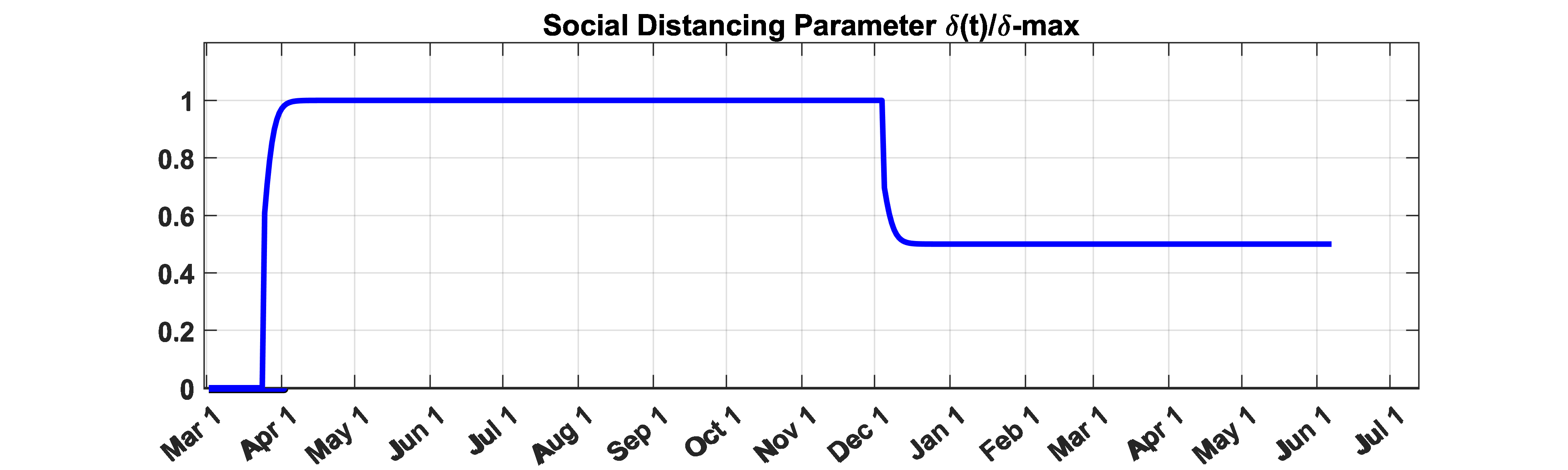}\\
      \includegraphics[width=0.475\textwidth]{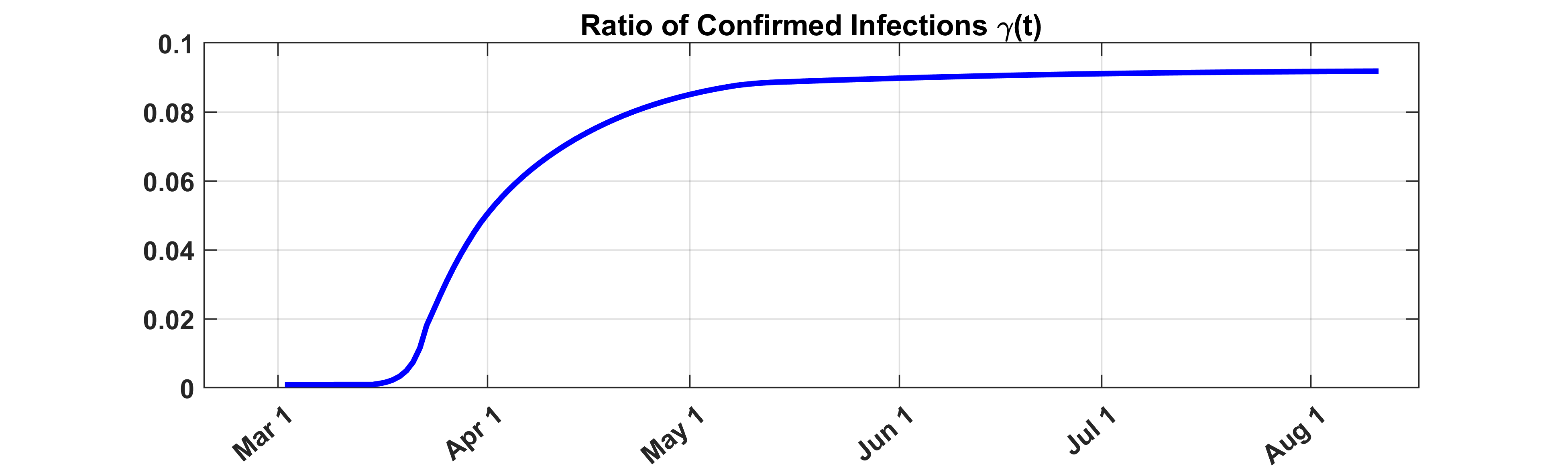}
    \end{tabular}\\
    (a) Scenario A.
    &
    (b) Scenario B.
  \end{tabular}
  \caption{{\bf New York (NY).}
    (Left) Scenario A relaxes control measures in accordance with real time policy decisions from each state \cite{JHU-Data,atlantic}. This results in relaxation of the lockdown measures on May 8, with social distancing relaxing around 80 days later and \math{\lambda_m} relaxing 43 days after the extension in social distancing. This is in accordance with the NY's current plan to delay school openings until September 23. 
  (Right) Scenario B extends social distancing and \math{\lambda_m} an additional 200 days after lockdown is relaxed, accounting for a scenario in which schools move to a remote learning model for the fall semester, and so the spike in infections is delayed until March 2021. In NY $R_{0}=6.24, \alpha_{max} =0.052, \delta_{max} = 0.051, \gamma_{today} = 0.09$, which corresponds to 1 in 11 infections being confirmed.}
  
\label{NYplots}
\end{figure}

\begin{figure}[H]
  \begin{tabular}{cc}
    \begin{tabular}{l}
      \includegraphics[width=0.475\textwidth]{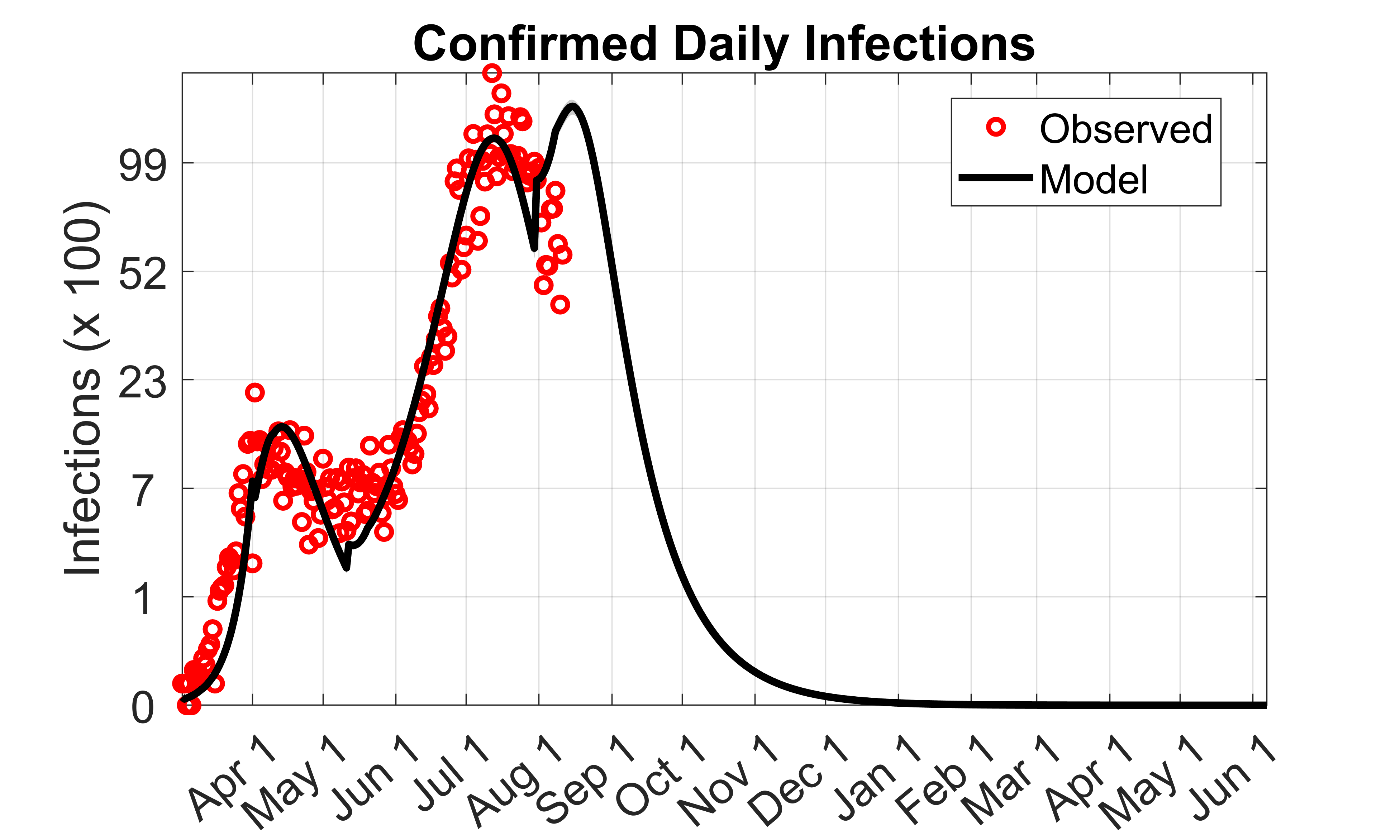}\\
      \includegraphics[width=0.475\textwidth]{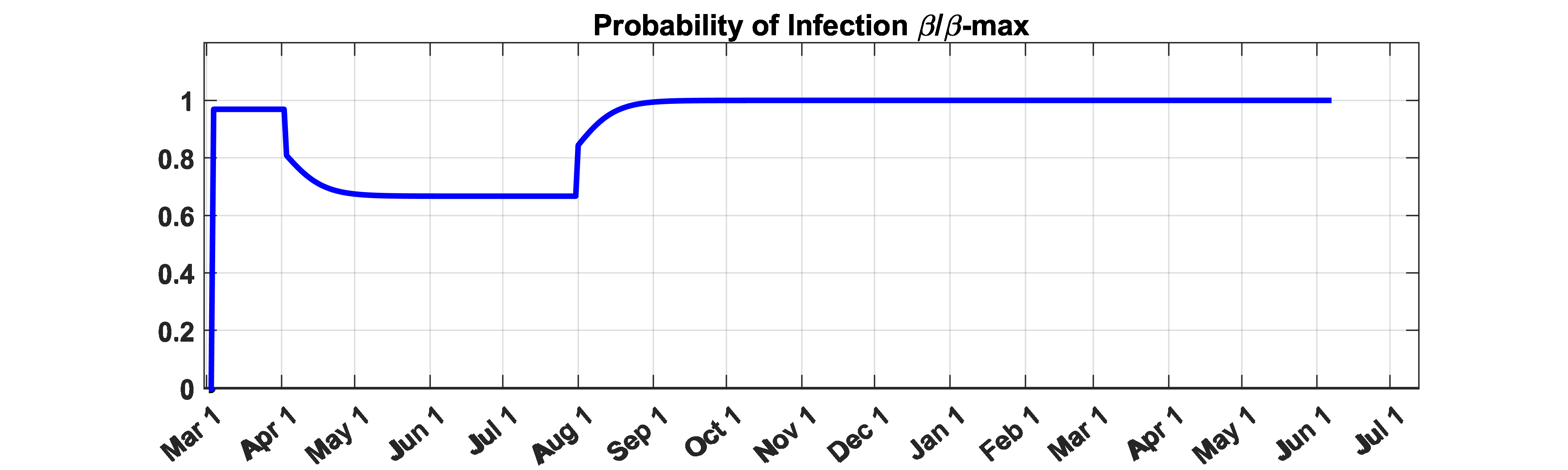}\\
      \includegraphics[width=0.475\textwidth]{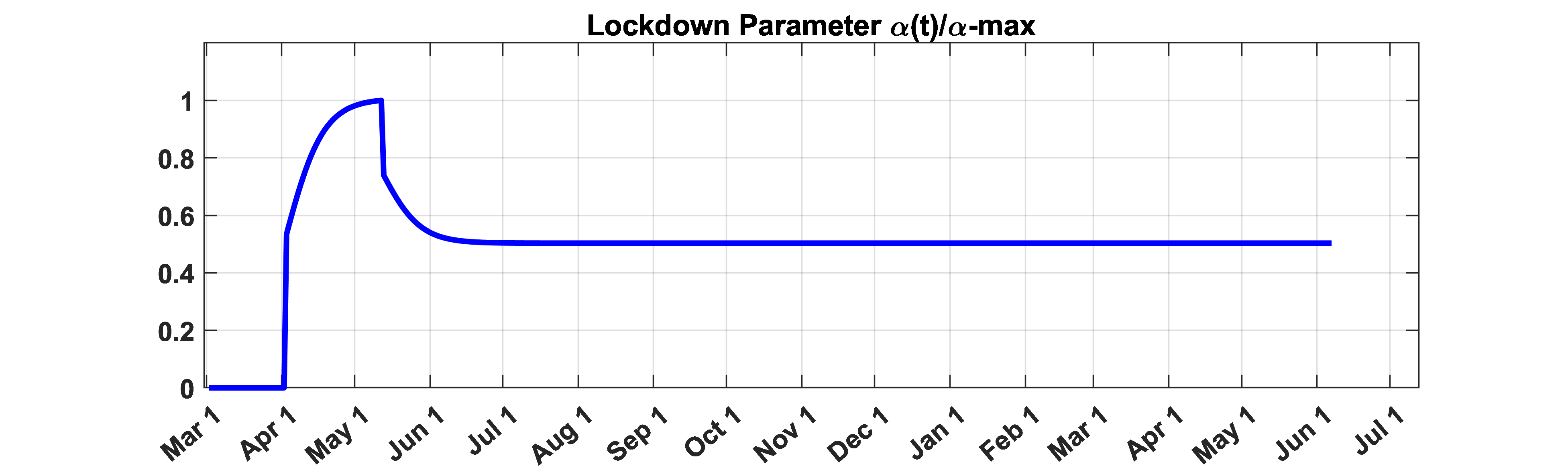}\\
      \includegraphics[width=0.475\textwidth]{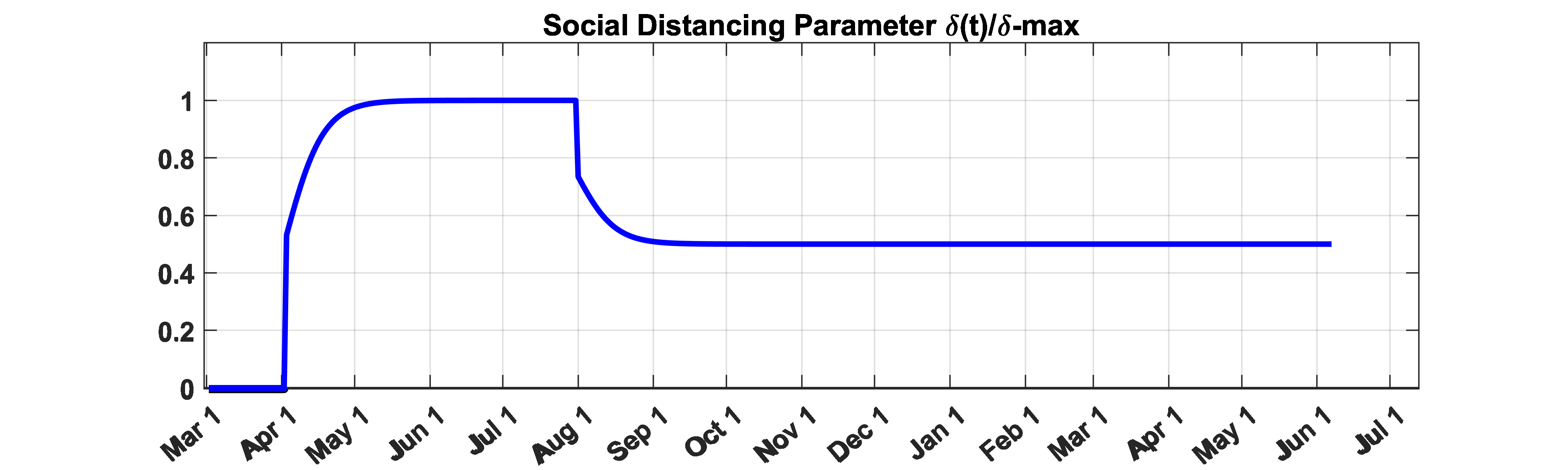}\\
      \includegraphics[width=0.475\textwidth]{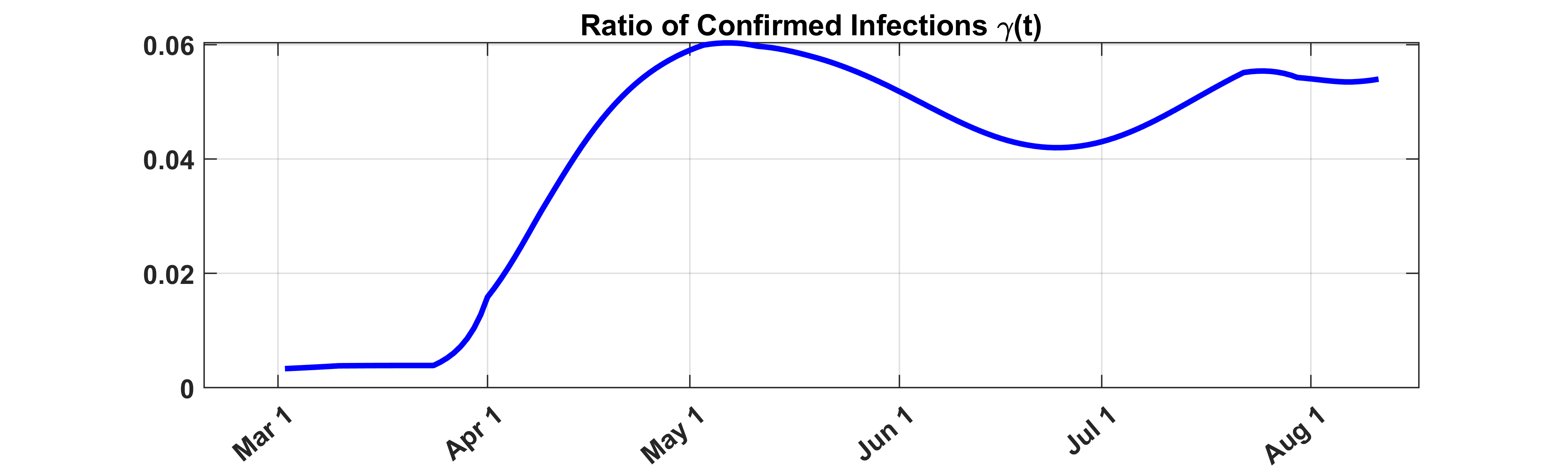}
    \end{tabular}
    &
    \begin{tabular}{l}
      \includegraphics[width=0.475\textwidth]{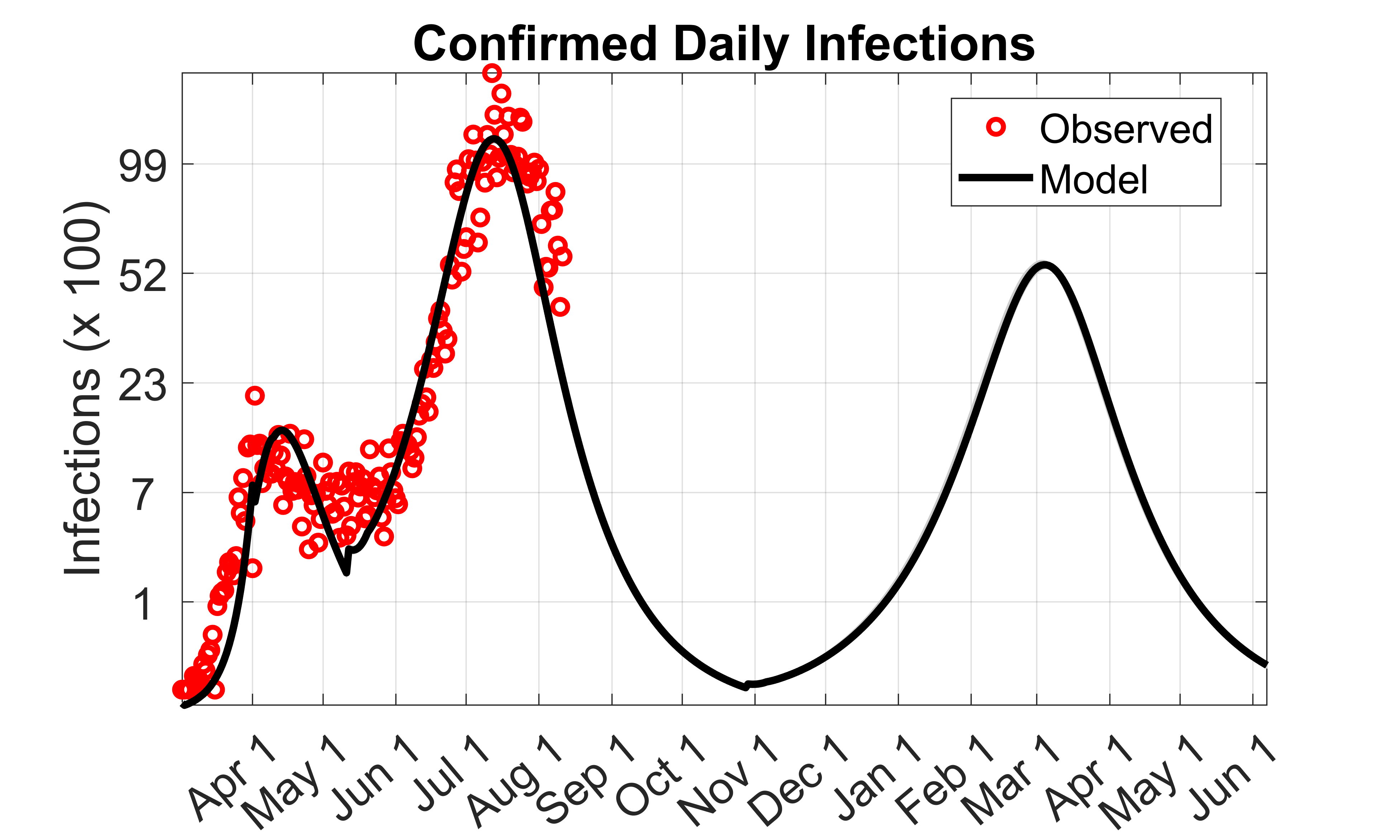}\\
      \includegraphics[width=0.475\textwidth]{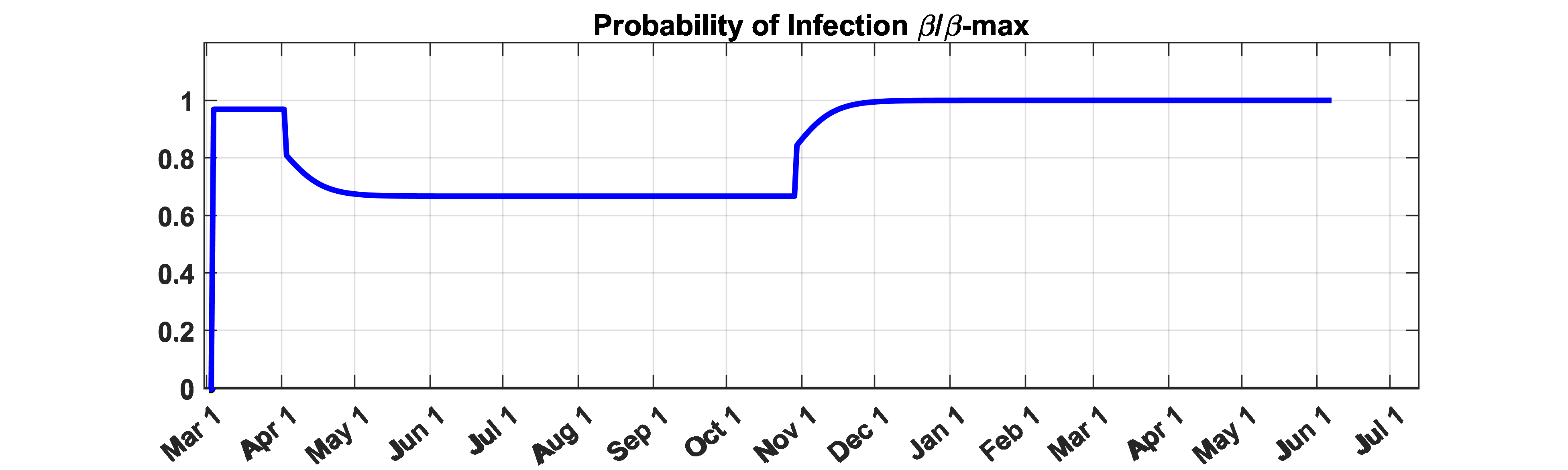}\\
      \includegraphics[width=0.475\textwidth]{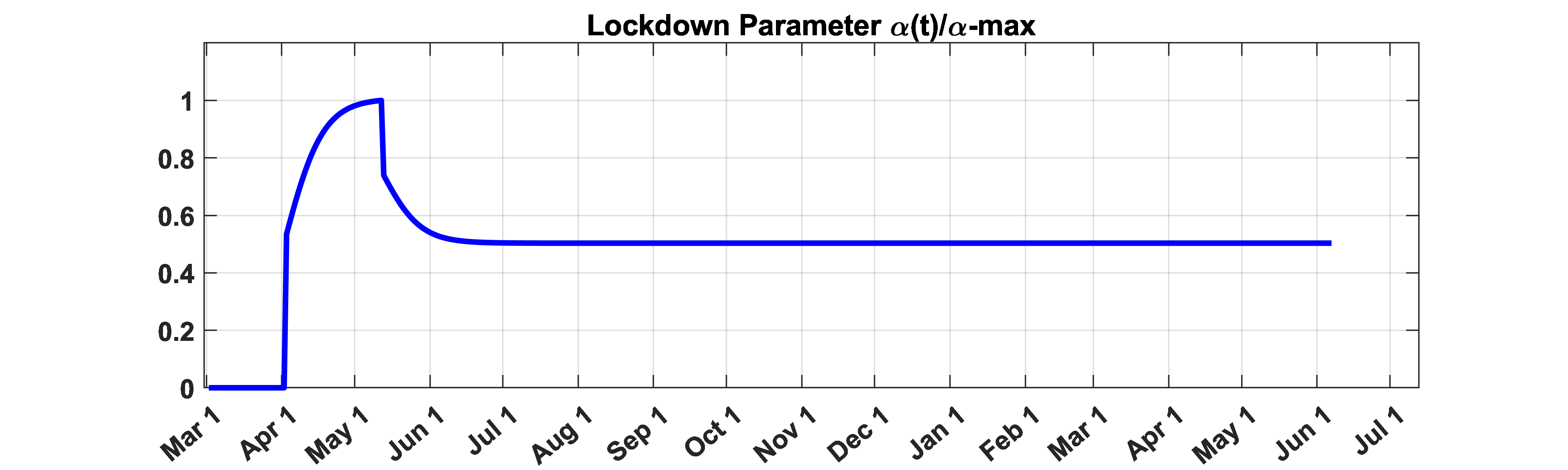}\\
      \includegraphics[width=0.475\textwidth]{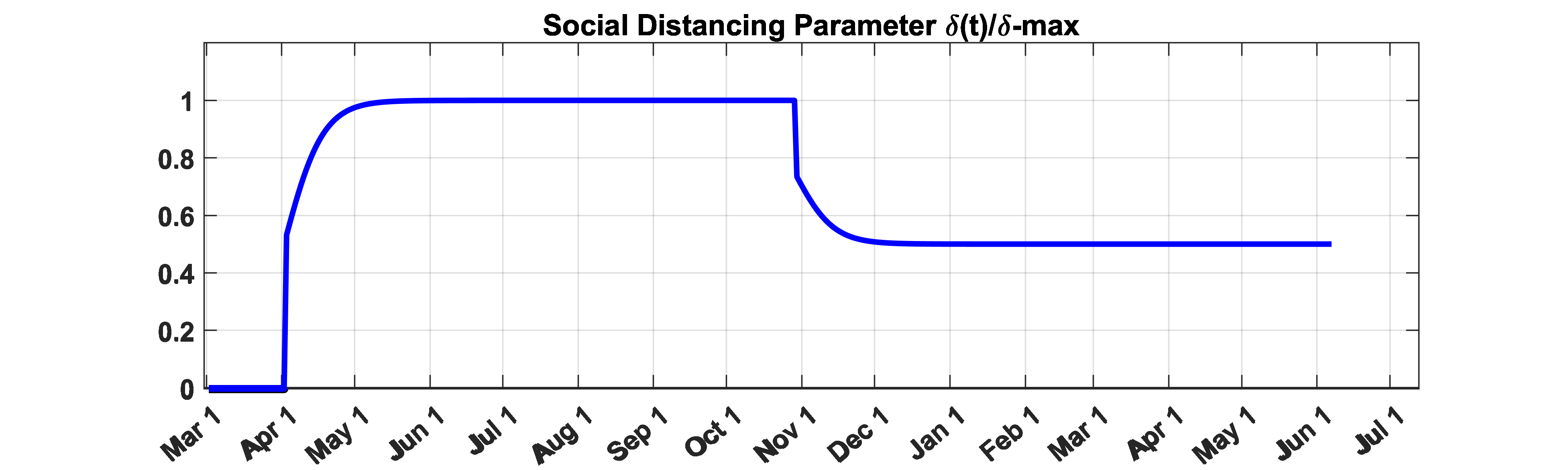}\\
      \includegraphics[width=0.475\textwidth]{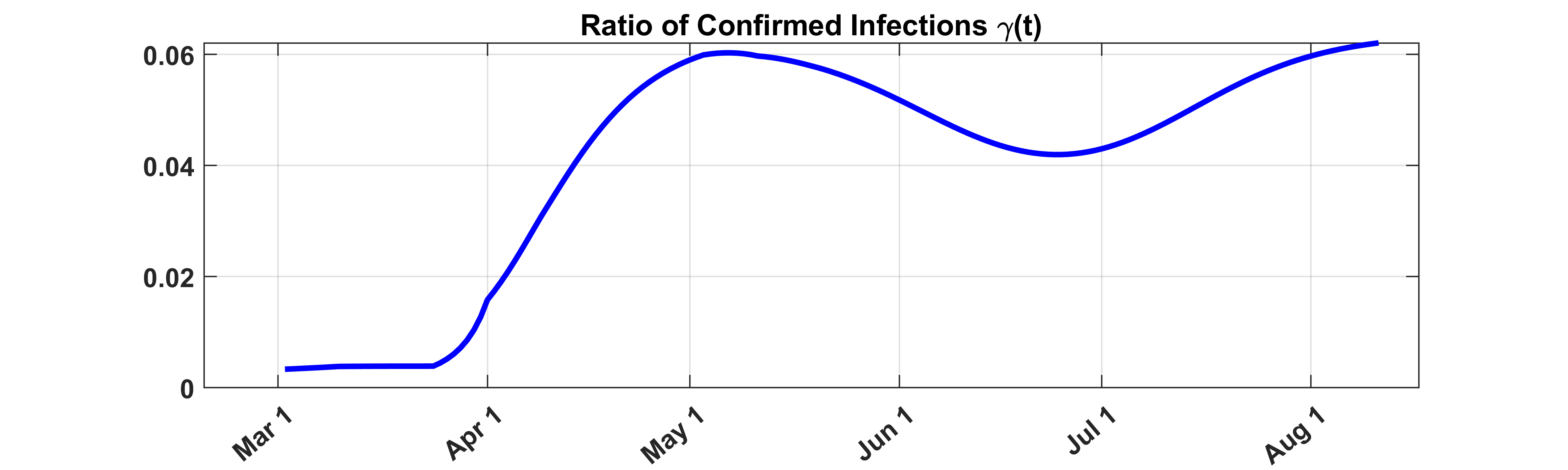}
    \end{tabular}\\
    (a) Scenario A.
    &
    (b) Scenario B.
  \end{tabular}
  \caption{{\bf Florida (FL).}
    (Left) Scenario A. In this case all control lockdown is relaxed around early May, while social distancing and \math{\lambda_m} measures are extended 80 days to the beginning of August and this seems to match the data in the confirmed infection curve very closely, predicting a spike in infections if social distancing restrictions are not extended past August
  (Right) Scenario B. In this case for Florida we extend social distancing and the \math{\lambda_m} measures for an additional 90 days in comparison to Scenario A, meaning social distancing and \math{\lambda_m} are now relaxed around early December. This not only pushes the second peak of back but reduces the peak height while increasing peak width, further spreading out infections. In  FL $R_{0} =3.56, \alpha_{max}=0.22, \delta_{max}=0.003, \gamma_{today} = 0.06$, which, similar to NC corresponds to approximately 1 in 15 infections being confirmed.
  For both of these scenarios we note the gray uncertainty bands are present but cannot be seen in the plots due to a relatively small value in the uncertainty.} 
  
\label{FLplots}
\end{figure}

The infection plots for the
three states are in Figure~\ref{NCplots} (North Carolina),
Figure~\ref{NYplots} (New York) and Figure~\ref{FLplots} (Florida).
We show the two scenarios, Scenario A and Scenario B. In each figure,
we first show the past
confirmed daily infections (red circles) from JHU data \cite{JHU-Data}
and the infection curves predicted by our new mathematical model (solid black line). Uncertainties in the infection curves are represented by gray uncertainty bands. To determine the uncertainty in the model predictions the algorithm was set to search for all models which have a fit-error within 5\% of the optimal. The gray uncertainty bands represent the set of possible outcomes from this set of models, and then the model (solid black line) is the average of these models. The largest source of uncertainty in our predictions is most likely due to noise in the publicly reported data. Because the machine learns from past data, any noise or uncertainty in that data is reflected as uncertainty in the future, and the farther the model projects into the future the larger the region of uncertainty.  
We then show the time dependent
learned coupling parameters \math{\alpha(t)},
\math{\delta(t)}, \math{\beta(t)}
(normalized to their maximum value), and \math{\gamma(t)},
see Equations \r{eq:lock1}, \r{eq:lock2}, \r{eq:lock3} and
\r{gamma}. Recall the definition of $\gamma$ as the reservoir parameter which in the variable definitions from Finite Difference model is computed
as \math{\gamma(t)=x(t)/(e(t)+x(t)+r_s(t))} from Equations \r{eq:diff3}-\r{eq:diff5}. In our illustration below,
we choose not to initiate a future full lockdown
as such a drastic  measure seems unlikely.
Instead, we focused on extending social distancing and mask mandates
which are more practical control measures. Our results show these are even more
effective in reducing infection curves than the lockdown.

We highlight several points from the figures.
\begin{enumerate}[noitemsep]
\item Each region and state has very different population densities and social interactions, hence the machine learning
  produces very different coupling parameters for each state. In contrast to SEIR model, this finding emphasizes
  the importance of space and time dependent restriction parameters contained in our model and the usefulness of machine learning in finding out this regional dependence. On our website~\cite{jmmmwebsite},
  we will show result to the county level.
\item The strength of the implemented
  control measures is determined by the level of social compliance.
  Since we do not know how compliant the population of a region is, it is necessary for
  the machine learning to automatically infer this from the observed past
  data during the phase when the virus spreading is completely random. This knowledge can be useful for other medical and viral diseases that rely on regional characteristics. The results vary considerably from county to county,
  once again underlining the value of heteregeonity.
\item Despite limited testing, we have here developed the tools to estimate the true number of non susceptible population, the 'reservoir', as a function of time in the past and for any predicted future scenario through parameter $\gamma(\vec{r},t)$ shown for the three states below. \\
\end{enumerate}

The area under the curve in all the plots shown gives the total confirmed number of infected, which through the parameter $\gamma$ of each region can be converted into a total number of non suscpetibles of each region.

\section{Conclusions}

The impact of the $Covid-19$ pandemic on countries around the world and on the daily life of citizens is unprecedented. The traditional class of models used by epidemiologists to study viral infections spreading randomly through a population has been the SEIR model we briefly reviewed in Section 1. The latter cannot account for the impact of global restriction measures on the pandemic, quantify the reservoir, or the space time dependence of these restrictions on the pandemic. We here offer a new and predictive mathematical model for the pandemic dynamics which combined with machine learning includes the space and time dependence in the pandemic dynamics.  
 
In contrast to previous viral diseases, the type and extent of global measures taken to contain the spread of $Covid-19$ to prevent a health crisis are new. In this new situation, part of the challenge is finding a predictive model for the evolution of the pandemic which describes the mix of order given by global restrictions, to the disorder of a random spreading of virus, and quantifies the impact of these control measures (their duration and the time they are introduced) in constraining the spread over extended periods of time. 

In this work we presented an interdisciplinary approach to predicting the dynamics of future infections. Firstly in Section 2 we develop a mathematical model which using physics based considerations to capture the effect of individual restrictions on the infection spread as a function of space and time. In our model the infection spread is a restricted self avoiding random walker through a population. Our model belongs in a different mathematical class from SEIR type models. It is stochastic and its probability distribution function $P(\vec{r}, t)$ on any region at any time is found by solving a Fokker Planck type equation (Eqn. \ref{fokkerplanck}). The strength of the lockdown and of social distancing and masks on the infection curves in our model are given by the time dependent parameters $\alpha(\vec{r}, t), \delta(\vec{r}, t)$ respectively. These restrictions introduce order in the random walk. The amount of disorder in the random walk which may persist through a population (for example a random mixing of susceptible and infectious groups through school openings) is still described by the $SEIR$ type term with strength $\beta(t)$, however note that in contrast to the $SEIR$ models, $\beta$ is a function of space and time and captures population demographics and social behavior. The diffusion of the infection through a population not captured in the above terms, (infections which somehow escape detection or the set of restrictions and quarantine) is given by a diffusion parameter $D(t)$ sourced by a noise term $f(t)$. This model can be applied to potential future pandemics, although of course we hope there wont be any.

Secondly, given the complexity of the model, we take full advantage of the benefits of AI in Section 3 in implementing our model in order to estimate future infection curves for any region within a short time. Viruses need hosts to feed  and multiply, therefore their unrestricted spreading phase depends on population demographics and human social behavior, in addition to the virus characteristic  reproductive rate $R_o$. We take advantage of this feature in training the machine. The machine learns the population density and social behavior networks for any region, state and country from the pre lockdown data  (March 2020) when the virus spread unconstrained. The machine then applies these learned characteristics when estimating future infections curves from our model and estimates the ratio of true infections for each confirmed infection $\gamma(\vec{r}, t)$. In this work we implemented a simplified version of our stochastic model by taking diffusion to zero  and solving the Langevin equation of Eqn. \ref{rateofinfection}. We deposited the code in Github \cite{ourcode} and will post examples of the predicted infection curves for each region located at $r$ as a function of a particular choice of restrictions $\alpha(t), \delta(t), \beta(t), \gamma(t)$ in our website \cite{jmmmwebsite}. 

For the case of time independent parameters, restricted SAW models in two dimensions can lead to critical phenomena of self similar hot and cold clusters of infections at large times. In reality, control measure parameters in our model are time dependent as restrictions change with time and country. Therefore, it is not clear whether the evolution of this pandemic mathematically would be in the same universality class, have the same fractal properties and polynomial growth at late times, as its well studied stationary SAW model. We plan to show our AI implemented infection curves for complete mathematical model presented here including diffusion in a sequel paper. Among other elements, diffusion is an important indicator of the possibility of re infection when travel is freely opened. In the present work we included the effect of border closing by imposing absorbing boundary conditions on our equations, meaning by taking the flux of population crossing boundaries in and out of a region to be zero.

How to interpret our results: Our new AI implemented model predicts future infection curves for any region as a function of the time and the restriction measures taken in that region. It explicitly quantifies the impact of each control measure on future infection curves over extended periods of time. It also quantifies the 'reservoir', through the parameter $\gamma$ which tells us the  number of undetected infections for each confirmed one in every region. Different scenarios, such as the choice of one restriction over another or a combination of them as well as the time they are imposed in the future, result in different predicted infection curves for the rest of the year. An added benefit is the speed of the machine in deriving these predictions.  We illustrate the comparison of different outcomes for the same region in Section 4 by showing the predicted infection curves for a few regions under two different scenarios, one with schools closed  and partial future lockdown and and social distancing rules, and one with schools open and no second lockdown but extended social distancing. The model can be applied to any pandemic or disease where the spreading of a disease is restrained by control measures. We hope that an advanced knowledge of the predicted infection curves as a function of regional restrictions given by our model will be useful in public health and economic considerations on the duration and combination of future restrictions which is optimal to the needs of their region.

\section*{Acknowledgements}
LMH would like to thank Dr. C. Carothers and Dr. B. Yener for their readiness to help with AI resources available at RPI, and Dr. S. Alexander and Dr. J. Engel for useful discussions in the early stage of this project. We thank Dr. D. Budenz  and Dr. D. Peterson for their time and expert advice on the public health and medical applications of our results. We are grateful to Dr. K. Simon and Dr. A. Bento fo their feedback on our work.
LMH acknowledges support from the Bahnson funds.

$\star$ We are saddened that one of our colleagues and a co-author in this paper, Steven Meshnick, passed away shortly before our paper was finished.



\bibliography{References}



\end{document}